\documentclass[lettersize,journal]{IEEEtran}
\usepackage{amsmath,amsfonts}
\usepackage{algorithmic}
\usepackage{algorithm}
\usepackage{array}
\usepackage[caption=false,font=normalsize,labelfont=sf,textfont=sf]{subfig}
\usepackage{textcomp}
\usepackage{stfloats}
\usepackage{url}
\usepackage{verbatim}
\usepackage{graphicx}
\usepackage{cite}
\graphicspath{{figs/}{figures/}{pictures/}{images/}{./}} % where to search for the images

\PassOptionsToPackage{hyphens}{url}
\usepackage{tabu}                      % only used for the table example
\usepackage{booktabs}                  % only used for the table example
\usepackage{lipsum}                    % used to generate placeholder text
\usepackage{mwe}                       % used to generate placeholder figures
\usepackage{ccicons}                   % package to be able to use icons from creative commons

\usepackage{mathptmx}                  % use matching math font
\usepackage{xcolor}
\usepackage{soul}
\usepackage{colortbl}
\usepackage{multirow}
\usepackage{booktabs, multirow, xcolor, tikz}
\usetikzlibrary{tikzmark, calc}
\usepackage[hidelinks]{hyperref}

\definecolor{circlecolor}{HTML}{7570B3}
\definecolor{ingestcolor}{HTML}{6a51a3}
\definecolor{detectcolor}{HTML}{238b45}
\definecolor{inspectcolor}{HTML}{1f78b4}
\definecolor{repaircolor}{HTML}{cb181d}
\newcommand{\roundedbox}[2]{%
  \tikz[baseline=(box.base)]{%
    \node[
      draw=#1, 
      fill=#1!20, 
      rounded corners,
      inner sep=1pt,
      font=\sffamily\small
    ] (box) {#2};
  }%
}

\newcommand{\ingest}[1]{\protect\roundedbox{ingestcolor}{#1}\xspace}
\newcommand{\detect}[1]{\protect\roundedbox{detectcolor}{#1}\xspace}
\newcommand{\inspect}[1]{\protect\roundedbox{inspectcolor}{#1}\xspace}
\newcommand{\repair}[1]{\protect\roundedbox{repaircolor}{#1}\xspace}

\newcommand{\tilen}{Tile2Net\xspace}

\newcommand{\system}{\textbf{Sidewalk Stewards}\xspace}

\newcommand{\myparagraph}[1]{\noindent\textbf{#1:}}

\usepackage{amssymb} %for the asterisk circle in thge user study figure

\newcommand{\highlight}[1]{#1}

\begin{document}

%% Paper title.
\title{Sidewalk Stewards:\\A Visual Analytics Approach for Interactive Topology Repair of Sidewalk Networks}

\author{Stefan Cobeli, Leonardo Ferreira, Logan Loch, Luiza Melo, Peyman Baghershahi,
        Claudio D. G. Linhares,\\Nivan Ferreira, Sourav Medya, Fabio Miranda,
        and~Maryam Hosseini%
\thanks{S. Cobeli, L. Ferreira, L. Loch, P. Baghershahi, S. Medya, and F. Miranda are with the University of Illinois Chicago. E-mail: scobel2@uic.edu.}%
\thanks{L. Melo and N. Ferreira are with the Universidade Federal de Pernambuco.}%
\thanks{C. D. G. Linhares is with Linnaeus University.}%
\thanks{M. Hosseini is with the University of California, Berkeley.}}

\maketitle

\begin{abstract}
Accurate pedestrian network data is foundational to urban accessibility, routing, and infrastructure planning, yet such data remains scarce and difficult to produce at scale.
Automated extraction pipelines have made significant strides by leveraging computer vision to derive sidewalk networks from aerial imagery, and have seen real-world adoption by municipalities seeking cost-effective inventory solutions.
Yet, these pipelines consistently produce topological errors (e.g., disconnected components, dangling edges, missed connections, malformed intersections) that undermine the utility of the extracted networks for routing and accessibility analysis.
Correcting these errors requires situated human judgment that global automated repair strategies cannot replicate, yet practitioners currently lack dedicated tools to support this work efficiently.
We present Sidewalk Stewards, a visual analytics system that bridges the gap between automated extraction and a deployment-ready pedestrian network.
Sidewalk Stewards detects candidate topology errors, supports interactive user-guided correction through coordinated spatial views, and provides real-time quality feedback as edits are applied.
\highlight{
We evaluate Sidewalk Stewards through a comparative user study, case studies across cities, and a quantitative evaluation.
Compared with conventional workflows, the system facilitates error identification and enables faster and more complete correction.
The case studies illustrate the workflow across different urban contexts, and the quantitative evaluation shows that a small number of user corrections can enhance multiple sidewalk quality metrics.
}
Sidewalk Stewards is available at \href{https://urbantk.org/sidewalk}{urbantk.org/sidewalk}.
\end{abstract}

\begin{IEEEkeywords}
Visual analytics, urban computing, sidewalks, built environment assessment, audit
\end{IEEEkeywords}

% \keywords{Visual analytics system, urban computing, sidewalks}

\section{Introduction}
\maketitle

% \IEEEPARstart{T}{his} 
\IEEEPARstart{S}{patial} geometric networks are integral to transportation, routing, and the everyday lives of people who rely on maps. For roads, large-scale digital mapping has been validated over time through public investment, federal and local efforts, and private companies such as Navteq and TomTom, and is continually refined through feedback from millions of GPS traces generated by drivers using navigation and mobility apps. 
For pedestrian infrastructure, there is no comparable ecosystem. We lack both the volume of trace data and the institutional infrastructure needed to validate topology, links, and network continuity at scale. 

Sidewalks are the arteries of pedestrian infrastructure and the backbone of pedestrian mobility, yet both the physical infrastructure and its digital representations are riddled with irregularities: gaps, discontinuities, and inconsistencies that compound one another and impact routing~\cite{rhoads2023sidewalk}, accessibility~\cite{deitz2021squeaky}, and planning alike~\cite{litman2023completing}.
These problems have direct, tangible consequences for the people who depend on them~\cite{cahen2025municipal}.
Routing engines that consume flawed sidewalk graphs return suboptimal or even dangerous paths~\cite{di2020walking}, steering pedestrians onto roads without crossings or into dead ends. Urban planners and transportation engineers who rely on network-level analysis to evaluate connectivity, identify service gaps, and prioritize infrastructure investments receive distorted pictures when the underlying topology is flawed. Safety researchers studying pedestrian exposure to vehicle traffic are similarly affected: a topologically broken network cannot faithfully represent the actual risk landscape walkers experience.

Much effort has been invested in the automated extraction of pedestrian networks~\cite{karimipedestrian}. These pipelines can ingest and process data at scale, and they have dramatically improved the baseline quality of available sidewalk graphs~\cite{hosseini2022walk}.
Among the most promising recent advances is \tilen~\cite{hosseini2023mapping}, a computer vision-based, end-to-end tool to automatically create sidewalk networks from high-resolution aerial imagery that has been successfully deployed in real-world urban mapping pipelines.
\tilen represents a significant step forward: it operates at the city scale, requires no manual digitization, and has demonstrated strong performance across diverse urban environments. Yet even state-of-the-art approaches consistently encounter a hard ceiling. Occlusions from tree canopies and building overhangs, ambiguous intersections, abrupt changes in pavement texture, and the sheer geometric variety of pedestrian infrastructure combine to produce residual segmentation errors that propagate into the extracted network topology. The resulting graphs suffer from disconnected components, spurious edges, and missed connection errors that statistical models and learned heuristics cannot fully self-correct. 
Closing this final gap is both time- and cost-intensive, as it requires human judgment, domain expertise, and contextual awareness that automated pipelines cannot replicate.

\highlight{
We relate these challenges to three problems of interest in visual analytics: guiding human inspection of automatically generated data using a multi-scale visual context, supporting data refinement through verification against contextual evidence, and incorporating user corrections into model refinement~\cite{xuan2025vista,bernard2018vial}.
However, these problems remain largely unaddressed for spatial data: prior visual analytics systems focus on exploratory analysis, with little attention to interactive editing and correction~\cite{https://doi.org/10.1111/cgf.14198}.
We examine these problems in the context of auditing errors in automatically extracted pedestrian networks and correcting their structure.
}
Addressing these challenges also stands to do more than improve map quality; it can meaningfully advance the civic goals that modern cities have committed to but struggle to operationalize.
Motivated by these challenges, we present \system, a visual analytics system designed to close that gap, directly tackling the topology errors and correction bottlenecks that automated pipelines like \tilen leave behind.
\system combines analytics with an interactive visual interface that makes errors legible and correctable.
The system surfaces candidate topology problems (disconnected components, geometric anomalies, accessibility inconsistencies), ranks them by estimated impact, and presents them through coordinated views that expose both local geometry and global network context. 
Analysts can inspect flagged locations, apply targeted repairs, and immediately observe how each correction propagates through downstream quality metrics, creating a feedback loop that supports efficient editing.
\highlight{This paper makes the following contributions: 
(1) A typology of sidewalk network errors that informs design goals and guidance for identifying candidate errors;
(2) A visual analytics tool that supports user-guided inspection and repair of detected errors through map interaction, structured editing operations, and guidance; 
(3) A repair model that adapts to the local context from the user’s corrections; 
(4) A set of case studies that demonstrate the system's utility in real-world scenarios;
(5) A comparative user study of finding and correcting sidewalk network errors with \system and a conventional GIS workflow using QGIS.}

% \maryam{This may come in handy: 
% One of the common problems of relying on road networks is the limited extent of the locations it covers. A network constructed based on streets and roads does not include any off-road footpath, pedestrian bridges, skywalks, or underground tunnels. In other words, it means being limited to only where roads can go, which can lead to inaccuracies (e.g., streets with no sidewalks), simplifications (e.g., assumptions that buildings can be directly accessed on both sides of a street centerline, while in reality crossing a street is only allowed at certain locations), and misrepresentation (e.g., assuming pedestrian connections based on vehicular routes, where there are none)~\cite{chin2008accessibility,ellis2016connectivity}, each of which can lead to potentially hazardous situations for pedestrians and those with specific needs~\cite{saha2019project}.
% }

% \fabio{Fabio will write}

% \begin{itemize}[noitemsep,topsep=0pt]
%     \item Motivation: pedestrian networks are messy in practice
%     \item Consequences for oruting, accessibility, urban planning, and safety
%     \item Limits of fully automatic extraction pipelines
%     \item Goal: enable users to efficiently identify and fix pedestrian network errors
%     \item Contributions: (1) Detects / prioritizes candidate pedestrian-network errors, (2) Supports interactive, user-guided correction workflows, (3) Provides quality feedback loops, (4) Case studies with practitioners
% \end{itemize}
\section{Related Work}

In order to position our contribution in the existing literature, this section reviews works in pedestrian network generation, spatial network visualization and visual analytics approaches to support data labeling, especially considering the use of active learning methods.

\subsection{Pedestrian network generation}

Prior approaches to sidewalk infrastructure mapping differ most fundamentally in how they establish the existence of pedestrian infrastructure: through direct observation, geometric inference, collective editing, or some combination thereof.
Pedestrian infrastructure can be mapped directly through field surveys, manual tracing from aerial imagery, or crowdsourced editing platforms~\cite{proulx2015database, manualtrace, walkscope}. These approaches can produce high-quality data, but they are labor intensive, difficult to scale, and often slow to update. 
Geometric buffering of road centerline data is another widely used method~\cite{karimipedestrian, brezina2017geometric}. This approach is easy to automate, but it relies on strong assumptions: that sidewalks exist along roads on both sides of the streets, and that crossings occur wherever roads intersect. As a result, it may generate pedestrian links where none exist and cannot represent off-road footpaths~\cite{chin2008accessibility,ellis2016connectivity,saha2019project}.
GPS trajectories have also been utilized to generate pedestrian networks~\cite{cottrill2013future,kasem2013pedestrian}. These methods capture where people actually walk rather than where infrastructure exists, but they depend on who is observed, where traces are dense enough, and how accurate the GPS data is ~\cite{yang2020pedestrian,karimipedestrian}.
LiDAR-based methods extract sidewalks and related features from point clouds using geometric, radiometric, or scan-pattern cues ~\cite{cheng2016extraction,lidarmass,balado2018automatic}. Since they capture three-dimensional structure directly, they can recover curbs, elevation changes, and path boundaries with high precision~\cite{horvath2022real,treccani2021sidewalk}. Their main limitations are uneven data availability and assumptions about surface form or material that may not generalize across cities~\cite{karimipedestrian,hosseini2022citysurfaces}.
Computer vision methods, and semantic segmentation in particular, are especially promising because they predict labels at the pixel level, enabling fine localization of narrow features such as sidewalks and crosswalks~\cite{ning2022sidewalk, hosseini2023mapping, ess2009segmentation, geiger2012we}. Automated pipelines like \tilen have demonstrated that city-scale pedestrian networks can be extracted from aerial imagery without manual intervention~\cite{hosseini2023mapping}. 
Yet automation does not eliminate error: segmentation models fail systematically on ambiguous surfaces, occluded paths, and features that lack training representation, and converting imperfect pixel predictions into usable network topology introduces further inconsistencies. 
The resulting networks require human review before they can support routing or accessibility analysis, but no tools exist to help analysts efficiently detect, understand, and correct errors at the scale these pipelines produce. \system aims to address this gap by supporting the entire pipeline for sidewalk network generation through a human-in-the-loop approach.

\subsection{Visual analytics and active learning for data labeling}

Pedestrian network generation can be understood as a data labeling problem: given imagery and derived representations, an analyst must attest to the presence or absence of sidewalk infrastructure at each location. We therefore survey visual analytics approaches designed to support annotation workflows.
Human-based labeling is inherently expensive, slow, and error-prone~\cite{bernard2018comparing}. Human-in-the-loop strategies that combine visual analytics with active learning can reduce these costs substantially across a range of data types~\cite{wang2011active, lu2017starpredictive}. In the text domain, Li et al.~\cite{li2025evovis} and Li et al.~\cite{li2026dall} couple data programming with active learning to support interactive labeling. In medical image segmentation, Yang et al.~\cite{yang2017suggestive} and Li et al.~\cite{li2023halia} propose active learning frameworks that reduce annotation burden while preserving model quality. More directly relevant to our setting, Xuan et al.'s VISTA system helps users identify, understand, and correct errors in foundation model-generated labels~\cite{xuan2025vista}, a goal structurally similar to ours, though targeting general-purpose vision models rather than geospatial network extraction.
Bernard et al. synthesized this body of work into a general framework called \emph{visual interactive labeling} (VIAL), which characterizes labeling pipelines that jointly produce labeled data, learned models, and analytical knowledge of both~\cite{bernard2018vial}. Our system instantiates this framework in the pedestrian network domain.
Similar ideas have been applied to urban labeling tasks. Kumar et al.~\cite{kumar2025geospatial} and Yu et al.~\cite{yu2023iterative} use active learning to support cool roof detection and road segment matching, respectively; Geiß et al.~\cite{geiss2017multitask} apply multi-strategy sample selection to classify building imagery across material, occupancy, and typology dimensions; and Islam et al.~\cite{islam2023active} develop an active learning framework for pedestrian and road sign detection in autonomous driving contexts.
Despite this breadth, a consistent limitation runs through these systems: the human participant functions as a passive labeler, responding to model queries without tools for proactively exploring data or diagnosing systematic error patterns. None targets the specific challenges of pedestrian network construction, where errors are spatially structured, topologically consequential, and difficult to detect without multi-scale visual context. \system addresses this gap by embedding active learning within a visual analytics environment designed to help analysts understand not just which examples to label, but also why the underlying model fails where it does.

% \sourav{these might help: road network , urban data \cite{,wu2024deep,geiss2017multitask}, pedestrian and road sign \cite{islam2023active},}

% \begin{itemize}[noitemsep,topsep=0pt]
%     \item Active learning has been used in visualization and visual analytics contexts~\cite{bernard2018comparing}
%     \item A quick summary of applications
%     \item try to find spatial/urban data examples
%     \item try to find network based examples
%     \item Ideally conclude that we have not seen a similar approach applied to the context of sidewalks
% \end{itemize}

\subsection{Visualization for spatial networks}

Spatial networks represent entities and relationships embedded in geographic space, such as transportation systems, pedestrian infrastructure, or mobility flows~\cite{Barth_lemy_2011}. 
Visual analytics has emerged as a powerful approach for exploring and analyzing such data by combining interactive visualization with computational analysis~\cite{Andrienko01092007, miranda2018shadow, moreira2023urban, moreira2024curio, miranda2024state, moreira2025urbanite}. Several studies in visualization have investigated how to effectively represent and analyze networks embedded in geographic space~\cite{https://doi.org/10.1111/cgf.14198,doi:10.1177/1473871612457601, srabanti2025streetweave}. Unlike abstract networks, spatial networks require integrating topological relationships with geographic constraints, creating challenges related to visual clutter, scale, and the integration of spatial context~\cite{Barth_lemy_2011}.
A comprehensive survey of geospatial network visualization techniques~\cite{https://doi.org/10.1111/cgf.14198} reviews methods for visualizing and interacting with networks embedded in geographic space. The survey characterizes existing approaches through a design space that considers the representation of geographic information, network structures, their visual integration, and interaction techniques. It highlights key trade-offs between preserving geographic fidelity and improving readability through transformations, such as layout adjustments or aggregation, while emphasizing the importance of interaction for tasks like pattern discovery, anomaly detection, and network comparison. Complementing this work, a meta-survey of network visualization research~\cite{https://doi.org/10.1111/cgf.14794} synthesizes existing surveys and task taxonomies and identifies geospatial network visualization as an increasingly important research direction. The analysis also outlines major challenges, including scalability, the integration of spatial and attribute information, and support for domain-specific analytical tasks.
Several visual analytics systems have been proposed to explore spatial or movement networks~\cite{https://doi.org/10.1111/cgf.14198, ferreira_assessing_2024}. For instance, systems for geospatial visual analytics combine maps with coordinated views to explore spatial patterns and relationships across multiple dimensions, enabling analysts to investigate spatial, temporal, and attribute-based patterns simultaneously. Early geovisual analytics systems, such as VIS-STAMP~\cite{visstamp}, integrate clustering, sorting, and coordinated visualizations to support the exploration of complex spatiotemporal datasets~\cite{doi:10.1177/1473871612457601}. More recent approaches focus on large-scale or complex movement networks, combining map-based visualizations with specialized views to reveal higher-order movement patterns and trajectories in urban environments~\cite{Feng2024}. These systems demonstrate the potential of visual analytics to support the exploration of spatial networks through interactive representations and multi-view analysis.
While many visual analytics systems focus on exploratory analysis, very little attention has been given to the interactive editing and correction of spatial data~\cite{https://doi.org/10.1111/cgf.14198}. 
Infrastructure datasets, such as road or pedestrian networks, often contain inconsistencies, missing connections, or topological errors introduced during data collection or automated extraction, which can negatively affect downstream analyses. 
We focus our efforts on this unexplored objective: not exploration of a given network, but auditing and correction of one that may be systematically wrong.

\section{Background}

% \begin{figure}
%   \centering\includegraphics[width=\linewidth]{example-image-a}
%   \caption{\fabio{Tile2net pipeline stages}}
%   \label{fig:tile2net}
% \end{figure}

\subsection{The pedestrian network data gap}

Despite sidewalks being among the most universally used urban infrastructures, comprehensive and accurate pedestrian network data remain remarkably scarce.
Unlike road networks, which have benefited from decades of investment by mapping agencies, automotive industries, and navigation companies, sidewalk infrastructure has no equivalent institutional champion~\cite{jane2022disparate}. 
National and municipal governments rarely maintain authoritative sidewalk inventories~\cite{evans2006sidewalk}, and where such records exist, they are often incomplete, outdated, or stored in formats incompatible with routing and analysis pipelines. 
OpenStreetMap, the most widely used open geographic database, offers global coverage but reflects the uneven enthusiasm of its volunteer contributor base: pedestrian network completeness varies dramatically across cities, neighborhoods, and income levels, with historically underserved communities consistently underrepresented~\cite{mooney2017review}. The result is a foundational data gap that undermines routing, accessibility planning, safety analysis, and infrastructure investment decisions at precisely the urban scales where these functions matter most~\cite{jane2022disparate}.

\subsection{Automated extraction with~\tilen}
\label{sec:tile2net}

To address this gap, researchers and practitioners have turned to automated extraction pipelines that derive pedestrian networks from remotely sensed data~\cite{karimipedestrian}.
Recent deep learning approaches have brought substantial advances, showing how neural networks trained on labeled overhead imagery could segment sidewalk surfaces, crosswalks, and footpaths with accuracy and generalizability~\cite{hosseini2023mapping}.
\tilen represents one strand of such models.
\tilen's pipeline begins with high-resolution aerial imagery, which is partitioned into fixed-size tiles and passed through a semantic segmentation model that classifies each pixel as belonging to one of several pedestrian infrastructure categories: sidewalk, crosswalk, footpath, or background. 
The resulting segmentation masks are then vectorized, i.e., converted from raster pixel regions into vector polygon geometries, and subsequently processed using a Voronoi-based approach to produce centerline representations of each infrastructure element. These centerlines are assembled into a graph structure in which nodes represent intersections and endpoints and edges represent traversable segments, yielding a routable pedestrian network derived entirely from overhead imagery without manual digitization. 
The pipeline is designed to operate at the city scale: given a bounding box or textual address and an imagery source,~\tilen can produce a draft pedestrian network for an entire metropolitan area in a matter of hours.
Crucially,~\tilen has moved beyond research prototype status. It has been adopted by municipal governments and transportation agencies seeking cost-effective methods to inventory and analyze their pedestrian infrastructure at city scale.
To date, the segmentation model hosted at ArcGIS Living Atlas has been downloaded over 39,000 times, signaling the pressing demand for automated pedestrian network data across urban governments with limited mapping budgets.

\subsection{Network creation and correction in practice}
% \maryam{I think we need to show that our method can work for any available network data that needs topology fix, it is important to show that our solution is not merely a tool built for another tool, it is addressing a larger problem. Showing one figure of the official municipal network data's network problems would be very helpful as well.}\fabio{Agree, can you write it?}

Operational experience with~\tilen has also exposed the limits of what automated extraction can achieve.
%
% \maryam{Adding the following and will get back to it later}
There are two main strands of errors in~\tilen: 1) Feature detection errors, such as misclassifications and missed detections; 2) Network generation errors resulting from sub-optimal interpolation distances in Voronoi-based centerline extraction. 
The output of~\tilen is thus best understood not as a finished product but as a high-quality draft: a major advance over manual tracing, but still requiring targeted human intervention before it can be trusted for routing, accessibility auditing, or policy analysis.
Practitioners who receive~\tilen outputs and need to produce deployment-ready pedestrian networks have developed ad hoc correction workflows that are laborious and difficult to scale. 
The typical process begins with a bulk quality assessment: loading the extracted network into a GIS environment such as QGIS or ArcGIS and visualizing the data to flag obvious errors.
Analysts then manually inspect flagged locations one by one, cross-referencing aerial imagery and local knowledge to determine whether a flag represents a genuine error and, if so, what the correct repair should be. 
Edits are made directly in the GIS attribute and geometry editors, with no structured support for prioritizing which errors to address first, no real-time feedback on how a repair affects downstream routing quality, and no systematic record of the decisions made. 
The process is expert-dependent, poorly reproducible, and offers no quality assurance mechanism beyond the individual analyst's judgment. For cities with thousands of kilometers of sidewalks to verify, this workflow is simply not viable at the scale that the pedestrian data gap demands.

% \begin{itemize}[noitemsep,topsep=0pt]
%     \item Lack of data related to pedestrian networks
%     \item Automated approaches are only part of the solution 
%     \item  \maryam{topology problems need local solutions, global fixes wont work (one size fits all solutions wont work)}
%     \item Network correction in practice, what is the usual workflow that experts employ to create these datasets
% \end{itemize}
\begin{figure}
  \centering\includegraphics[width=\linewidth]{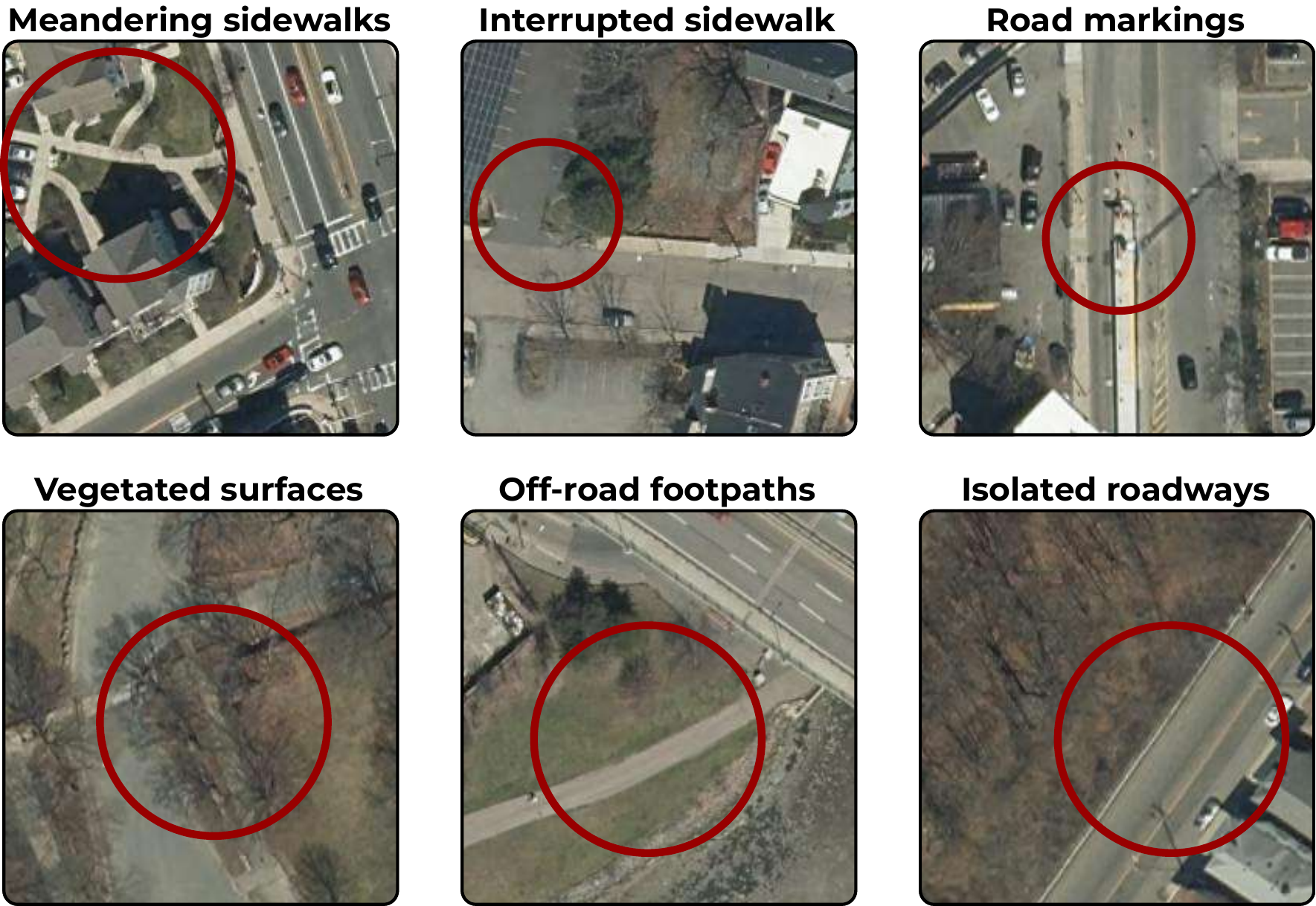}
  \caption{Typology of common errors with \tilen; the circled areas highlight the types of regions that are commonly misclassified.}
  \label{fig:typology}
\end{figure}

\section{From Network Errors to Design Goals}

% \fabio{Fabio should write the intro (also change the title of the section to something better)}
% \maryam{updated this part}

% \maryam{Again, I think this needs to be more generic and then we show~\tilen as a case. We provide our solution as one addressing a general strand of problems, instead of only Tile2nNet problems. The framing should change. Added my version here.}

Geospatial networks are only as useful as their topology. A broken link or misaligned node propagates directly into routing errors and accessibility miscalculations. Automated extraction pipelines have reduced the cost of generating pedestrian network data at the city scale, but their outputs contain recurring errors that limit direct use for routing, accessibility analysis, and planning. These are not model-specific failures; they reflect a broader class of problems encountered by researchers, municipal data producers, and practitioners who produce and validate such data.
We organize these problems into three families. First, \textbf{representation errors} arise from how pedestrian space is modeled. Unlike roads, there is no single standard representation for sidewalks, footpaths, open spaces, and plazas. The appropriate representation is case-based and driven by downstream use: whether an open space should be encoded as a boundary polygon, a diagonal connector, or a local link between adjacent segments depends on how that space will be traversed and analyzed.
Second, \textbf{detection errors} arise when segmentation models miss valid pedestrian infrastructure or assign pedestrian classes to non-pedestrian surfaces.
Third, \textbf{topology and conversion errors} arise when detected features are converted into a network, introducing broken links, dangling segments, gaps at intersections, misaligned nodes, or invalid crossings.
We use~\tilen as a case through which to examine these problems. Close inspection of its outputs surfaces failure modes across all three families. We then derive a set of design goals for \system that address the mentioned class of problems rather than the limitations of a single tool.

\subsection{Typology of errors}

Errors in automated pedestrian network extraction are not uniformly distributed; they tend to concentrate around specific configurations of the built environment. Characterizing these configurations is a prerequisite for building a system that can prioritize inspection and support active learning.  Both capabilities require that errors have predictable, learnable structures rather than occurring at random. The typology below is our empirical evidence that such a structure exists.
We carried out a systematic visual inspection of a dataset comprising 1,000 tiles from Boston. We chose Boston due to the unique challenges its urban fabric poses: a high-density urban area with an organic grid of narrow, winding streets and complex intersections, tall buildings in the downtown core that produce highly shadowed and visually obstructed sidewalks, paired with dense vegetation cover, piers, trails, and pedestrian-dedicated spaces along the waterfront and park corridors. The model was not trained on Boston, making it a suitable case for out-of-distribution analysis.
These tiles were overlaid with corresponding~\tilen predictions and ground truth polygons. 
Segmentation errors were examined per tile, enabling observation of the conditions under which they occurred. 
This inspection produced a preliminary listing of approximately eighteen urban elements associated with recurring segmentation errors. A second stage involved randomly selecting 500 tiles for detailed analysis, which enabled the refinement of the grouping criteria: related cases were merged and non-recurrent instances were excluded, yielding a consistent set of six representative types.
To complement the visual inspection, numerical descriptors from~\tilen were computed for each tile using the prediction and confidence masks, including uncertainty statistics, confidence distributions, sidewalk pixel counts, and revised recall metrics. 
These metrics were aggregated by typology category to assess whether the visually defined classes correspond to consistent statistical patterns.

The categories below emerge from the intersection of visual inspection and quantitative validation: each represents a morphological configuration that was both visually recurrent across the dataset and associated with a consistently elevated error rate, defined as the proportion of inspected tiles in which that configuration co-occurred with a segmentation error.
Figure~\ref{fig:typology} presents all typology classes.
Visual inspection repeatedly surfaced errors in areas with irregular or curved sidewalk geometries, a pattern confirmed by an error rate of 29.51\%. These errors do not originate from the detection stage but arise during the network construction phase. Extracting clean centerlines from elongated or geometrically complex polygons is difficult, and sub-optimal interpolation distances could introduce artifacts in the resulting graph, a known challenge that the~\tilen authors discuss directly. We identify this configuration as \textbf{meandering sidewalks}.
Disconnected segments appearing precisely where driveways and alleys cross sidewalks emerged as one of the most visually consistent patterns in the dataset,  with 38.35\% of inspected tiles exhibiting this error. This is likely a training data artifact: planimetric maps, which formed the basis of the capture rules and annotation schema, treat alleys and driveways as interruptions in the sidewalk layer, and sidewalks were therefore systematically not mapped at those locations. The model learned this convention, producing disconnections that reflect the annotation standard rather than the physical reality of the infrastructure. We identify this configuration as \textbf{interrupted sidewalks}.
Footpaths in vegetated areas repeatedly went undetected during inspection, with a 37.16\% error rate. Dense canopy and strong shadows obscure paved surfaces, making them difficult to distinguish from surrounding vegetation. We identify this configuration as \textbf{off-road footpaths and trails}.
Roads in low-density suburban areas were frequently misclassified as sidewalk, a pattern observed in 36.72\% of inspected tiles. The likely mechanism is a training data artifact: the model was trained predominantly on high-density urban environments where narrow paths through vegetated areas are consistently park footpaths, annotated as sidewalk. The model learned this association, which breaks down in low-density suburban contexts where similarly narrow, vegetation-flanked features are local roads rather than pedestrian infrastructure. We identify this configuration as \textbf{isolated roadways}.

Boundary errors in areas with dense vegetation recurred consistently across the dataset, with an error rate of 47.04\% across inspected tiles. Vegetation along sidewalk edges obscures the transition between paved and unpaved surfaces, making boundary resolution the primary failure mode rather than interior detection. We identify this configuration as \textbf{vegetated surfaces}.
Lane markings and painted road patterns were repeatedly misclassified as crosswalks, introducing spurious edges into the extracted network, with an error rate of 24.98\% across inspected tiles. We identify this configuration as \textbf{road markings}.

\subsection{Design goals and requirements}

The typology above establishes that~\tilen errors are not random: they concentrate in identifiable spatial configurations and propagate from pixel-level misclassifications into network-level faults in predictable ways. This structure is what makes systematic triage and iterative refinement possible, and it directly motivates the design goals below. 
To understand how these errors manifest in practice, we also collected informal feedback over several months from GIS analysts and transportation planners who had directly used~\tilen outputs in production mapping workflows. 
These conversations emerged organically from ongoing collaborations with practitioners who had encountered~\tilen's limitations firsthand and had developed ad hoc correction strategies in response. 
Across these exchanges, a consistent set of pain points recurred: no structured means of prioritizing which errors to address, no real-time feedback on how a repair affects downstream network quality, and no safeguards against inadvertently introducing new faults while fixing existing ones. 
The design goals below are derived directly from these observed failure modes and workflow bottlenecks.
Each goal reflects a specific gap between what automated extraction produces and what a deployment-ready pedestrian network requires.

\noindent \textbf{DG1: Prioritize candidate errors.}
Practitioners spend a disproportionate share of their time locating errors rather than correcting them, relying on visual scanning of large network exports with no systematic ordering.
The typology confirms that errors concentrate in identifiable spatial configurations: configurations like \textbf{open spaces and plazas} and \textbf{vegetated surfaces}, which carry high failure rates, should rise to the top of the inspection queue.
The system should automatically surface these candidates and rank them so that analysts can direct attention to the most consequential faults first.

\noindent \textbf{DG2: Provide evidence to support repair decisions.}
Many error categories are inherently ambiguous from overhead imagery alone. Whether a surface adjacent to a road is a valid sidewalk or a parking lot in the case of \textbf{isolated roadway surfaces}, or whether a sidewalk actually exists beneath a dense tree canopy in the case of \textbf{vegetated surfaces} and \textbf{off-road footpaths and trails}, are questions that aerial imagery cannot reliably resolve. 
Practitioners frequently switch between multiple tools (e.g., aerial imagery viewers, and ground-level street imagery) to resolve many of these ambiguities and to gather enough evidence to make a confident repair decision. The system should present this evidence in coordinated views so that context is immediately available without switching applications.

\noindent \textbf{DG3: Support navigation across scales and editing operations.}
Sidewalk network errors span multiple levels of granularity. 
For example, \textbf{Meandering sidewalks} and \textbf{interrupted sidewalks} only become fully legible at the micro scale of individual edges and intersection geometries.
At the macro scale, aggregate uncertainty metrics allow analysts to identify which regions are most in need of attention.
The system should support fluid navigation from a macro view of network-wide uncertainty down to the micro level of a specific error, so that analysts can understand both the global context of a fault and the local geometry needed to correct it.

\noindent \textbf{DG4: Support validation and refinement through active learning.}
\textbf{Isolated roadway surfaces} and \textbf{interrupted sidewalks} both reflect consistent, learnable biases; a model that systematically misclassifies suburban road margins or loses continuity at driveways is exhibiting a correctable pattern rather than random noise. 
Each analyst correction on these configurations is a training signal. The system should close the loop between repair actions and model behavior, supporting an active learning workflow in which analyst decisions progressively recalibrate the model toward the morphological characteristics of the city being corrected.

\begin{figure*}[t!]
  \centering
  \includegraphics[width=\textwidth]{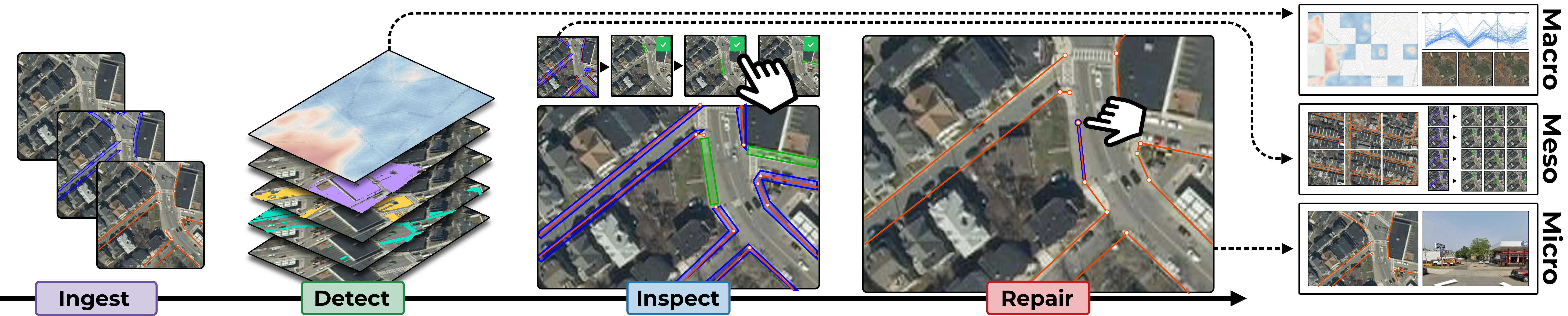}
  \caption{\system is organized along two dimensions: a data pipeline (horizontal axis) spanning ingestion to network repair, and a scale framework (vertical axis) that structures detection, inspection, and correction across three spatial scales. The pipeline stages advance from raw imagery through uncertainty estimation, interactive inspection, and topology editing; the scale framework ensures that each pipeline stage can be engaged by the user at the granularity appropriate to the task, from city-wide triage down to individual segment correction.}
    \vspace{-0.5cm}
  \label{fig:workflow}
\end{figure*}

\noindent \textbf{DG5: Make uncertainty explicit.}
Not all error categories carry the same interpretive confidence. \textbf{Off-road footpaths and trails} and \textbf{vegetated surfaces} are cases where occlusion makes it genuinely unclear whether pedestrian infrastructure exists at all.
Treating all flagged locations with equal confidence makes it difficult for analysts to calibrate how much verification effort each candidate warrants. 
The system should surface uncertainty explicitly so that triage decisions are better informed.
\noindent \textbf{DG6: Scalability to city-scale networks.}
Pedestrian networks for even a single city can encompass thousands of kilometers of sidewalk and tens of thousands of individual edges, making it impractical to load, inspect, or correct the full network at once. 
The system should remain responsive and usable at this scale, supporting efficient data loading, rendering, and error detection without requiring analysts to work on artificially restricted subsets of the network.

\noindent \textbf{DG7: Support for incremental workflows.}
Correcting a city-scale network is not a task that can be completed in a single session. Analysts may work across multiple sittings, hand off work to colleagues, or return to previously reviewed areas as new errors are discovered. The system should preserve the state of prior corrections so that work is not lost.

\newcommand{\tablebox}[3]{%
  \tikz[baseline=(box.base)]{%
    \node[
      draw=#1,
      fill=#1!20,
      rounded corners,
      inner sep=2pt,
      font=\small\sffamily,
      text width=#2,
      align=center
    ] (box) {#3};
  }%
}

% \begin{table}[t]
% \centering
% \caption{System features across pipeline stages and analytical scales.}
% \label{tab:features}
% \setlength{\tabcolsep}{4pt}
% \renewcommand{\arraystretch}{0}
% \begin{tabular}{@{}l
%                 >{\centering\arraybackslash}p{2.0cm}
%                 >{\centering\arraybackslash}p{2.0cm}
%                 >{\centering\arraybackslash}p{2.0cm}@{}}
% \toprule
% \textbf{Stage}
%   & \textbf{Macro} & \textbf{Meso} & \textbf{Micro} \\
% \midrule

% Ingest
%   & \multicolumn{3}{c}{\tablebox{ingestcolor}{6.2cm}{Load tiles \& graph}} \\
% \midrule

% \multirow{2}{*}{Detect}
%   & \multicolumn{2}{c}{\tablebox{detectcolor}{3.9cm}{Uncertainty computation}} & \\
%   & \multicolumn{2}{c}{\tablebox{detectcolor}{3.9cm}{SAM masks}} & \\
% \midrule

% \multirow{2}{*}{Inspect}
%   & \multicolumn{2}{c}{\tablebox{inspectcolor}{3.9cm}{Coordinated views}} & \\
%   & \multicolumn{2}{c}{\tablebox{inspectcolor}{3.9cm}{Tile ranking}} & \\
% \midrule

% \multirow{3}{*}{Repair}
%   & & \multicolumn{2}{c}{\tablebox{repaircolor}{3.9cm}{Repair suggestions}} \\
%   & & \multicolumn{2}{c}{\tablebox{repaircolor}{3.9cm}{Repair model}} \\
%     & & & \multicolumn{1}{c}{\tablebox{repaircolor}{1.65cm}{Direct editing}} \\
% \bottomrule
% \end{tabular}
% \label{table:features}
% \end{table}

\begin{table}[t]
\centering
\highlight{%
\caption{System features across pipeline stages and analytical scales,
with the design goals they support.}
\label{table:features}
\setlength{\tabcolsep}{4pt}
\renewcommand{\arraystretch}{1}
\resizebox{\columnwidth}{!}{%
\begin{tabular}{@{}l
                >{\centering\arraybackslash}p{2.0cm}
                >{\centering\arraybackslash}p{2.0cm}
                >{\centering\arraybackslash}p{2.0cm}
                >{\centering\arraybackslash}p{2.0cm}@{}}
\toprule
\textbf{Stage}
  & \textbf{Macro} & \textbf{Meso} & \textbf{Micro}
  & \textbf{Design goals} \\
\midrule

Ingest
  & \multicolumn{3}{c}{\tablebox{ingestcolor}{6.2cm}{Load tiles \& graph}}
  & DG6 \\
\midrule

\multirow[t]{2}{*}{Detect}
  & \multicolumn{2}{c}{\tablebox{detectcolor}{3.9cm}{Uncertainty computation}}
  & & DG1, DG5 \\
  & \multicolumn{2}{c}{\tablebox{detectcolor}{3.9cm}{SAM masks}}
  & & DG1 \\
\midrule

\multirow[t]{2}{*}{Inspect}
  & \multicolumn{2}{c}{\tablebox{inspectcolor}{3.9cm}{Coordinated views}}
  & & DG1, DG2, DG3, DG5 \\
  & \multicolumn{2}{c}{\tablebox{inspectcolor}{3.9cm}{Tile ranking}}
  & & DG1 \\
\midrule

\multirow[t]{3}{*}{Repair}
  & & \multicolumn{2}{c}{\tablebox{repaircolor}{3.9cm}{Repair suggestions}}
  & DG3, DG4 \\
  & & \multicolumn{2}{c}{\tablebox{repaircolor}{3.9cm}{Repair model}}
  & DG4 \\
  & & & \tablebox{repaircolor}{1.65cm}{Direct editing}
  & DG3, DG4 \\
\bottomrule
\end{tabular}%
}
}
\end{table}

\section{\system}

\subsection{Overview}

\system is a human-in-the-loop visual analytics tool designed to bridge the gap between~\tilen's automated extraction output and a clean, deployable pedestrian network. 
The system is organized along two dimensions: a \textbf{data pipeline} that structures \emph{what} the system does at each stage of the correction process, and a \textbf{scale framework} that structures \emph{where} the analyst engages with the network. Together, these two dimensions define both the system's architecture and the analyst's workflow.
Across these dimensions, \system is composed of a set of modules, shown in Table~\ref{table:features}, specifically designed to tackle the previous design goals.
Before detailing \system' main components, we next give an overview of both dimensions.

\subsubsection{\system's data pipeline}

The data pipeline guides the analyst through a structured $ingest\rightarrow detect\rightarrow inspect \rightarrow repair$ cycle.
In the \textbf{ingestion stage}, \system loads two complementary  representations of the same urban area: the orthorectified aerial tiles and the extracted pedestrian graph produced by~\tilen. The result of this step is a spatially coherent workspace in which graph elements can be examined against the imagery from which they were derived.
In the \textbf{detection stage}, a Segment Anything Model (SAM) \cite{carion2025sam} is applied to the aerial tiles to produce semantic masks of built environment features (e.g., pavement surfaces, building footprints, vegetation canopy). These masks, together with an \textbf{uncertainty} computation, characterize the spatial distribution of potential errors at the macro and meso scales, directing the analyst's attention towards the most problematic areas of the network.
In the \textbf{inspection stage}, the analyst explores candidate errors through \textbf{coordinated views} that juxtapose graph geometry and aerial imagery at the macro and meso scales, and a \textbf{ranking} mechanism that orders candidates by estimated severity. To reduce the cognitive burden of repair, the system surfaces suggested fixes derived from~\tilen's own output. The analyst can accept, modify, or reject each suggestion.
In the \textbf{repair stage}, the analyst applies corrections at the meso and micro scales. A \textbf{suggestions} panel surfaces~\tilen-derived repair candidates in context; a \textbf{repair model} trained on accumulated analyst decisions learns to anticipate corrections from graph and imagery features, progressively reducing manual effort; and \textbf{network editing} tools allow direct manipulation of individual edges and nodes at the micro scale. Each confirmed repair is recorded in a provenance log that captures the corrective action applied, providing a training signal for the repair model.
This cycle is iterative: repairs may resolve clusters of errors at once or reveal previously masked issues, and the analyst may move freely between inspection and repair until the network meets their quality threshold.

\subsubsection{\system's scale framework}

The macro-meso-micro framework organizes the analyst's spatial interaction with the network across three nested scales, from city-level overview to individual element. 
At the \textbf{macro scale}, the analyst surveys the entire city to identify where errors are most concentrated and decides where to focus effort. Interactions here are primarily observational, reading aggregate quality signals rather than editing individual elements. 
Zooming in, the \textbf{meso scale} shows a neighborhood or corridor in context. Here, the analyst consults the ranking to determine which candidates to address first and reviews suggestions before committing to a repair strategy.
At the finest granularity, the \textbf{micro scale} presents small sections of the network with enough local context (aerial imagery, exact geometry) to apply a repair. Here, the analyst cross-references aerial imagery and street-level imagery to confirm the nature of an error, evaluate suggested fixes, and apply a direct edit to the network using the editing tools.
The three scales are tightly coordinated. Selecting an area of interest at the macro scale zooms the meso view to the relevant area; selecting a candidate error at the meso scale opens the micro inspector. Conversely, a repair applied at the micro scale propagates quality signals upward, updating the repair model.

% \begin{itemize}[noitemsep,topsep=0pt]
%     \item Pipeline
%     \item Data model (entities, graph representation, derived features?)
%     \item Error taxonomy: connectivity errors, geometric errors, semantic errors, redundancy errors
%     \item Introduce macro, meso, micro framework.
% \end{itemize}

\subsection{Ingestion \& detection: Preprocessing sidewalk data}

Before analysis can begin, raw aerial imagery must be ingested and processed into a  network. The preprocessing straddles the ingestion and detection stages and encompasses three main components: \ingest{loading}, \detect{heatmap}, and \detect{SAM masks}.

\myparagraph{Ingestion}
The \ingest{loading} component accepts as input the georeferenced tile imagery and, optionally, the corresponding network output produced by~\tilen.
If no output is provided, \system will run~\tilen by itself and create the corresponding network output.
\tilen operates in three stages described in Section~\ref{sec:tile2net}.

\myparagraph{Detection}
Once ingested, the tiles are passed through two complementary detection components.
The first \detect{uncertainty} component aggregates~\tilen's per-pixel segmentation uncertainty scores into a tile-level average, producing a heatmap that reveals which areas of the city the model found most ambiguous.
Tiles with high average uncertainty are candidates for containing extraction errors, as they might correspond to locations where the segmentation model lacked confidence (e.g., areas that have occlusion, shadows, or atypical pavement configurations).
This heatmap provides the macro- and meso-scale signal that can guide analysts towards the regions most in need of review.
The \detect{SAM masks} component runs SAM on each tile, generating masks that delineate surface boundaries independently of~\tilen's classifications based on semantically meaningful categories drawn from the typology of errors. These categories include vegetation, buildings, driveways, and tree canopy.
For instance, since the taxonomy revealed that connectivity and geometric errors are  concentrated around the tree canopy, where occlusion causes~\tilen to fragment or misplace sidewalk segments, the tree canopy class becomes a particularly informative feature for flagging candidate errors in a given tile.

\subsection{Inspection: Guiding users towards errors}

The inspection stage is where analysts decide which candidates to prioritize. This stage is organized around \inspect{coordinated views} and \inspect{tile ranks} operating at the macro and meso scales of the framework.

The \inspect{coordinated views} are composed of a map view, parallel coordinates plot (PCP), and a tile gallery. The map view renders the full extent of the processed area, encoding each tile's uncertainty or SAM category distributions as a heatmap. Analysts can toggle between these distributions to support reading both the geographic structure of model confidence and environmental factors.
In another view, each tile is represented as a polyline in a PCP with axes corresponding to~\tilen's average uncertainty score and the coverage fractions of each SAM category. This view shows the distribution of features across all tiles to support the isolation of clusters of tiles that share a particular profile (e.g., tiles with high uncertainty and high tree canopy coverage) by brushing across one or more axes. 
Finally, the \inspect{tile ranks} display small images of tiles ranked by whichever PCP dimension the analyst has selected. As the analyst interacts with the PCP, the gallery updates in real time to surface the tiles most representative of the selected feature range.
Selecting a tile in the gallery highlights its corresponding polyline in the PCP and centers the map on its location, closing the coordination loop across all three views.

\subsection{Repair: Topology correction and model retraining}

Once an analyst has identified a candidate error for correction, the repair stage supports both direct editing and 
% model-assisted suggestion 
\highlight{repair suggestions}
and feeds the resulting decisions back into the repair model through a retraining loop. There are three components in this stage: a component that creates \repair{repair suggestions}, a \repair{repair model} that is trained on user-validated fixes, and interactions to support \repair{direct editing}.

For each element the analyst decides to correct, \system generates correction candidates by analyzing the network topology in the vicinity of the error. 
% Specifically, \system identifies dangling endpoints (nodes with only one connected edge) and proposes connections to the nearest compatible endpoints within a spatial tolerance.
\highlight{
Specifically, \system identifies dangling endpoints (nodes with only one connected edge) and considers pairs separated by less than 60 meters.
Endpoints within a boundary margin equal to 1\% of the tile's smaller dimension are excluded, and each endpoint must lie inside or within 0.5 meters of an input polygon.
For each eligible pair, the system generates a rectangular connection and clips it to the tile boundary.
To discourage suggestions that cross streets diagonally, we use a road-overlap heuristic: suggestions are discarded when 50--75\% of their area overlaps road polygons extracted from \tilen's predictions.
}
These \repair{repair suggestions} are rendered as overlays in the tile gallery so that analysts can accept the most geometrically plausible repair, reject all suggestions, and draw a correction manually, or adjust a suggestion. 
% This selection-based interaction minimizes time-consuming annotation effortg
\highlight{This selection-based interaction is intended to reduce annotation effort}
 while producing unambiguous data points, since each accepted repair constitutes a clean (input, correction) pair for downstream training.
Before accepting or adjusting a suggestion, analysts can verify the proposed repair against ground-level evidence using the embedded Google Street View panel, which renders a street-level panorama for the selected location. This is useful for ambiguous cases, such as paths obscured by tree canopy or driveways that visually resemble sidewalks, where aerial imagery alone is insufficient to determine whether a connection should exist.
When no candidate repair is satisfactory, analysts can fall back to a \repair{direct editing} mode that operates at the micro scale.  The editing interface supports node dragging, edge insertion, and segment deletion directly on top of the aerial imagery. All editing operations support a snapping mechanism that aligns moved or newly drawn endpoints to the nearest existing node.

Finally, \system contains a \repair{repair model} trained on accumulated corrections, whether accepted from suggestions or applied through direct editing. The model uses a U-Net segmentation architecture~\cite{ronneberger2015u} with a ResNet34 encoder~\cite{he2016deep} pretrained on ImageNet~\cite{deng2009imagenet}. The encoder is frozen except for the first convolutional layer, which is modified to accept a 5-channel input: three RGB channels from the satellite tile, a fourth channel for the \tilen rasterized polygon mask, and a fifth channel for a \tilen confidence map derived from the softmax output of its segmentation network. The decoder is fully trainable. Training targets the segmentation regions missed by \tilen, with labels computed as the difference between the ground truth mask and the \tilen polygon mask. Data augmentation includes horizontal and vertical flips, random 90-degree rotations, and random crops applied to all channels; color jitter and Gaussian noise are applied to the RGB channels only to preserve the integrity of the confidence and polygon masks. We use focal loss~\cite{lin2017focal}, which down-weights correctly classified pixels and concentrates learning on misclassified ones.
For the problematic regions, the repair model replaces the segmentation part, taking the aerial imagery tiles as input and learning to produce corrected segmentation masks that reflect the fixes analysts have validated downstream.
This was a deliberate design choice: segmentation is the stage where environmental confounders such as tree canopy, shadows, and driveway surfaces introduce the most systematic errors, and it is the stage whose outputs propagate directly into all subsequent~\tilen's vectorization, centerline generation, and graph assembly steps.

\subsection{Implementation}

\begin{figure}[t]
  \centering\includegraphics[width=\linewidth]{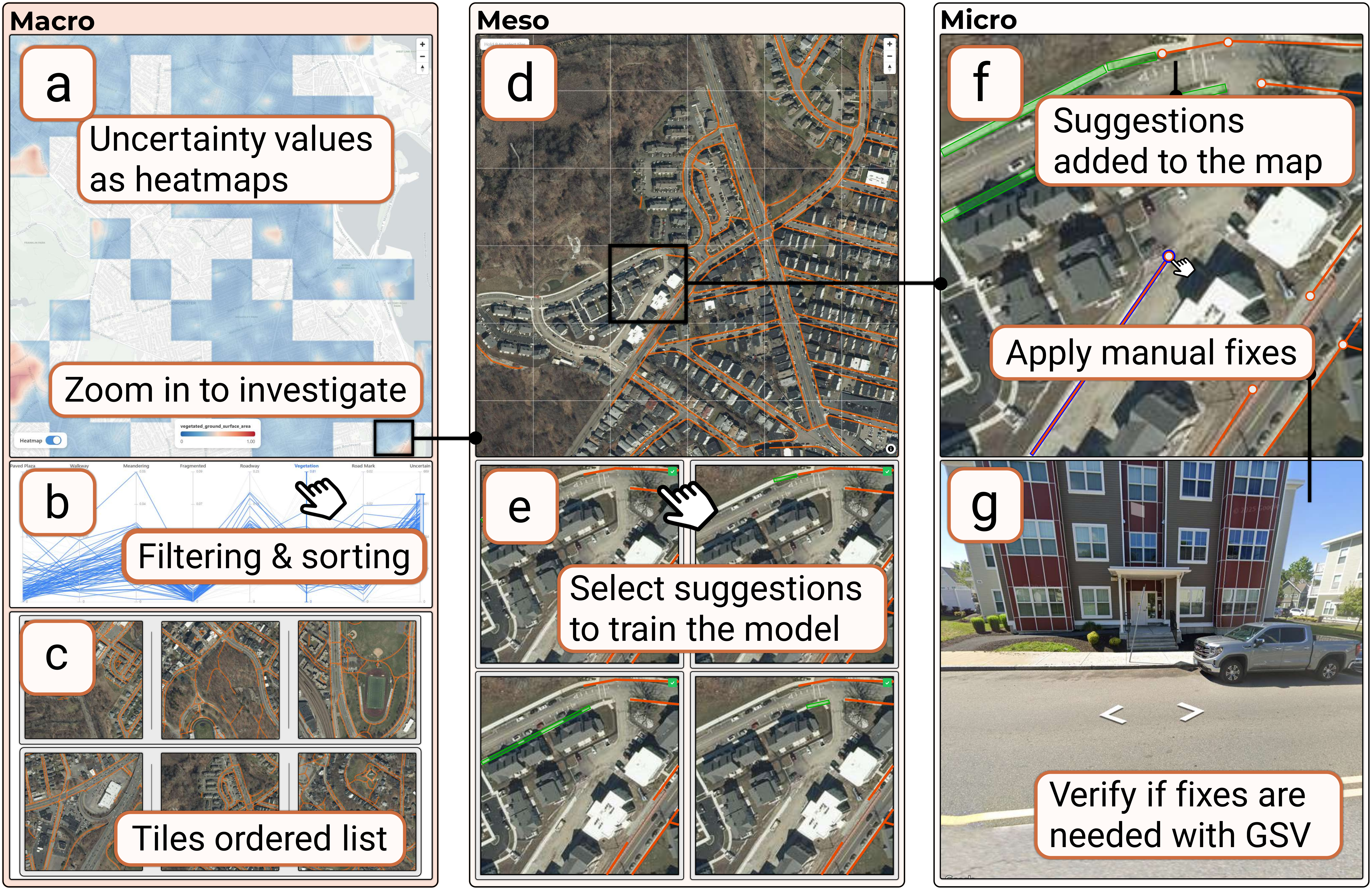}
  \caption{Main visual components of the \system system.}
  \label{fig:interface}
\end{figure}

% \fabio{Leonardo should write this}

% \begin{itemize}[noitemsep,topsep=0pt]
%     \item three-component architecture: tile service for imagery, served as a raster tile pyramid (python), backend API for uncertainty computation/training/inference (python), and VA web interface (react).
%     \item load~\tilen outputs (graph + polygons) + orthorectified tiles as aligned workspace for analysis (shared workspace where imagery and network geometry can be inspected together).
%     \item uncertainty + segmentation-based feature masks (e.g., vegetation) exposed as a heatmap at macro scale.
%     \item coordinated views (map, pcp, and tile ranking) support filtering, sorting, and tile comparison, enabling navigation from city-scale views to localized regions.
%     \item system supports model-driven suggestions (polygon candidates), by allowing users to select direct suggestions from the mode, suggestion creation, by allowing users to directly draw/modify suggestions on the map, and direct editing nodes and edges for correcting topology.
%     \item user corrections/suggestions are collected and sent to the backend to retrain the model and generate predictions (human-in-the-loop refinement), that are then, rendered on the interface.
% \end{itemize}

\system is built on a three-component architecture: a Python tile service that delivers orthorectified aerial imagery as a raster tile pyramid, a Python backend API responsible for uncertainty computation, model training, and inference, and a React-based visual analytics web interface. 
The system loads \tilen outputs (the extracted network graph and associated polygons) alongside the aligned imagery tiles into a shared workspace where network geometry and source imagery can be inspected together. At the macro scale, uncertainty scores and segmentation-based feature masks (such as vegetation coverage) are visualized as a heatmap, giving users an immediate overview of where errors are likely concentrated. 
Figure~\ref{fig:interface} highlights the main components of the \system interface.
A set of coordinated views (the map (a), a parallel coordinates plot (b), and a tile ranking gallery (c)) support filtering, sorting, and comparison, enabling fluid navigation from city-scale overviews down to localized regions of interest. 
For repair, the system supports three modes of interaction: accepting 
% model-driven polygon suggestions 
\highlight{repair suggestions} (d,e), creating new suggestions by drawing or modifying geometries directly on the map (f), and editing nodes and edges to correct network topology. 
The user can at any time visualize street-level imagery (g).
User corrections and new suggestions are collected and sent to the backend, where they retrain the  model and generate updated predictions that are rendered back in the interface, closing a human-in-the-loop refinement cycle.
\section{Evaluation}

\begin{table*}[t!]
\centering
\highlight{
\caption{Comparing \tilen , \system' repair model, and the ground-truth networks on the held-out validation tiles: standard and revised segmentation IoU and Rec, number of connected components per tile, number of dangling endpoints (nodes with degree 1), proportion of reachable pairs, and Average Path Length Similarity (APLS). Deltas are $|$t2n$-$GT$|-|$repair$-$GT$|$: positive (teal) means the repair moved closer to the ground truth.}
\label{tab:network_evaluation}
}
\vspace{-0.25cm}
% \resizebox{0.48\textwidth}{!}{\begin{tabular}{llllllllll}
\resizebox{\textwidth}{!}{\begin{tabular}{llllllllll}
\toprule
\textbf{Difficulty} & \textbf{Model} & \textbf{rIoU} & \textbf{rRec} & \textbf{IoU} & \textbf{Rec} & \textbf{\#Components} & \textbf{\#Dangling} & \textbf{Reachability} & \textbf{APLS} \\
\midrule
\multirow{6}{*}{[0, 0.25)} & t2n & 0.03 & 0.05 & 0.01 & 0.02 & 1.03 & 0.62 & 0.03 & 0.03 \\
 & repair-5 & 0.15 {\color{teal}(+0.12)} & 0.32 {\color{teal}(+0.27)} & 0.06 {\color{teal}(+0.05)} & 0.15 {\color{teal}(+0.13)} & 1.95 {\color{teal}(+0.81)} & 2.08 {\color{purple}(-0.27)} & 0.08 {\color{teal}(+0.05)} & 0.08 {\color{teal}(+0.05)} \\
 & repair-10 & 0.21 {\color{teal}(+0.18)} & 0.61 {\color{teal}(+0.56)} & 0.09 {\color{teal}(+0.08)} & 0.34 {\color{teal}(+0.32)} & 2.70 {\color{teal}(+0.05)} & 2.81 {\color{purple}(-1.00)} & 0.14 {\color{teal}(+0.11)} & 0.14 {\color{teal}(+0.11)} \\
 & repair-20 & 0.22 {\color{teal}(+0.19)} & 0.51 {\color{teal}(+0.46)} & 0.08 {\color{teal}(+0.07)} & 0.26 {\color{teal}(+0.24)} & 2.86 {\color{purple}(-0.11)} & 2.97 {\color{purple}(-1.16)} & 0.08 {\color{teal}(+0.05)} & 0.07 {\color{teal}(+0.04)} \\
 & repair-recife-20 & 0.03 (0) & 0.05 (0) & 0.01 (0) & 0.02 (0) & 1.05 {\color{teal}(+0.03)} & 0.59 {\color{purple}(-0.03)} & 0.03 (0) & 0.03 (0) \\
 & GT & 1.00 & 1.00 & 1.00 & 1.00 & 1.89 & 1.22 & 1.00 & 1.00 \\
\midrule
\multirow{6}{*}{[0.25, 0.5)} & t2n & 0.27 & 0.39 & 0.12 & 0.18 & 2.37 & 1.16 & 0.23 & 0.22 \\
 & repair-5 & 0.37 {\color{teal}(+0.10)} & 0.54 {\color{teal}(+0.15)} & 0.16 {\color{teal}(+0.04)} & 0.27 {\color{teal}(+0.09)} & 3.00 {\color{teal}(+0.63)} & 2.32 {\color{purple}(-0.84)} & 0.23 (0) & 0.22 {\color{purple}(-0.01)} \\
 & repair-10 & 0.48 {\color{teal}(+0.20)} & 0.77 {\color{teal}(+0.39)} & 0.23 {\color{teal}(+0.11)} & 0.52 {\color{teal}(+0.34)} & 3.21 {\color{teal}(+0.84)} & 2.84 {\color{purple}(-1.37)} & 0.39 {\color{teal}(+0.16)} & 0.33 {\color{teal}(+0.11)} \\
 & repair-20 & 0.50 {\color{teal}(+0.22)} & 0.70 {\color{teal}(+0.31)} & 0.22 {\color{teal}(+0.10)} & 0.42 {\color{teal}(+0.24)} & 3.58 {\color{teal}(+1.00)} & 2.74 {\color{purple}(-1.26)} & 0.44 {\color{teal}(+0.21)} & 0.40 {\color{teal}(+0.18)} \\
 & repair-recife-20 & 0.28 (0) & 0.39 (0) & 0.12 (0) & 0.18 (0) & 2.37 (0) & 1.16 (0) & 0.23 (0) & 0.22 (0) \\
 & GT & 1.00 & 1.00 & 1.00 & 1.00 & 3.47 & 1.32 & 1.00 & 1.00 \\
\midrule
\multirow{6}{*}{[0.5, 0.75)} & t2n & 0.55 & 0.64 & 0.26 & 0.33 & 2.28 & 1.53 & 0.30 & 0.29 \\
 & repair-5 & 0.58 {\color{teal}(+0.02)} & 0.73 {\color{teal}(+0.09)} & 0.28 {\color{teal}(+0.02)} & 0.42 {\color{teal}(+0.08)} & 2.67 {\color{teal}(+0.40)} & 1.91 {\color{purple}(-0.37)} & 0.27 {\color{purple}(-0.03)} & 0.26 {\color{purple}(-0.04)} \\
 & repair-10 & 0.61 {\color{teal}(+0.06)} & 0.77 {\color{teal}(+0.13)} & 0.29 {\color{teal}(+0.03)} & 0.46 {\color{teal}(+0.13)} & 3.33 {\color{teal}(+1.05)} & 2.77 {\color{purple}(-1.23)} & 0.27 {\color{purple}(-0.03)} & 0.25 {\color{purple}(-0.04)} \\
 & repair-20 & 0.64 {\color{teal}(+0.09)} & 0.77 {\color{teal}(+0.14)} & 0.30 {\color{teal}(+0.04)} & 0.45 {\color{teal}(+0.12)} & 3.63 {\color{teal}(+0.98)} & 3.00 {\color{purple}(-1.47)} & 0.43 {\color{teal}(+0.13)} & 0.42 {\color{teal}(+0.13)} \\
 & repair-recife-20 & 0.55 (0) & 0.64 (0) & 0.26 (0) & 0.33 (0) & 2.28 (0) & 1.53 (0) & 0.32 {\color{teal}(+0.02)} & 0.31 {\color{teal}(+0.02)} \\
 & GT & 1.00 & 1.00 & 1.00 & 1.00 & 3.44 & 1.44 & 1.00 & 1.00 \\
\midrule
\multirow{6}{*}{[0.75, 1.0]} & t2n & 0.91 & 0.98 & 0.56 & 0.77 & 2.97 & 0.71 & 0.74 & 0.72 \\
 & repair-5 & 0.89 {\color{purple}(-0.02)} & 0.99 (0) & 0.54 {\color{purple}(-0.01)} & 0.78 {\color{teal}(+0.01)} & 3.03 {\color{teal}(+0.06)} & 0.90 {\color{teal}(+0.20)} & 0.71 {\color{purple}(-0.03)} & 0.69 {\color{purple}(-0.03)} \\
 & repair-10 & 0.85 {\color{purple}(-0.06)} & 0.99 {\color{teal}(+0.01)} & 0.52 {\color{purple}(-0.04)} & 0.79 {\color{teal}(+0.02)} & 3.19 {\color{purple}(-0.05)} & 1.33 (0) & 0.68 {\color{purple}(-0.06)} & 0.66 {\color{purple}(-0.06)} \\
 & repair-20 & 0.88 {\color{purple}(-0.03)} & 0.99 {\color{teal}(+0.01)} & 0.54 {\color{purple}(-0.02)} & 0.79 {\color{teal}(+0.02)} & 3.32 {\color{purple}(-0.17)} & 1.24 {\color{teal}(+0.09)} & 0.76 {\color{teal}(+0.02)} & 0.74 {\color{teal}(+0.02)} \\
 & repair-recife-20 & 0.91 (0) & 0.98 (0) & 0.54 {\color{purple}(-0.01)} & 0.77 (0) & 3.00 {\color{teal}(+0.04)} & 0.71 (0) & 0.73 {\color{purple}(-0.02)} & 0.70 {\color{purple}(-0.01)} \\
 & GT & 1.00 & 1.00 & 1.00 & 1.00 & 3.06 & 1.02 & 1.00 & 1.00 \\
\bottomrule
\end{tabular}}
\vspace{-0.5cm}
\end{table*}

% \system was evaluated through three case studies and structured feedback sessions with five urban planners, none of whom are authors of this paper.

\highlight{
We evaluated \system from multiple perspectives: quantitatively assessing the repair model (Section~\ref{subsec:quantitative_evaluation}), comparing the end-to-end correction workflow against a conventional GIS workflow in a user study (Section~\ref{subsec:user_study}), examining its use across three real-world case studies (Section~\ref{subsec:case_studies}), and gathering structured feedback from domain experts (Section~\ref{subsec:expert_feedback}).
}

%\system was evaluated through a comparative user study, three case studies, a quantitative evaluation of the repair model, and structured feedback sessions.}

%The structured feedback sessions were conducted with five urban planners, none of whom are authors of this paper.

\highlight{\subsection{Quantitative evaluation} \label{subsec:quantitative_evaluation}}

\highlight{To quantify whether corrections collected through \system can improve pedestrian network extraction, we compared the standard use of~\tilen against variants augmented with a learned repair model.
Our baseline applied~\tilen directly using its pre-trained weights, reflecting the conventional workflow. The remaining configurations represent the workflow enabled by \system: user corrections made through the interface are used to fine-tune a repair model that refines \tilen's segmentation output.
Because the repair model modifies only the segmentation stage, while leaving vectorization and network generation unchanged, improvements can propagate through the existing pipeline to the resulting pedestrian network.
We evaluated four repair-model configurations: three trained with 5, 10, and 20 corrections collected during a typical \system session in Boston, and one trained with 20 corrections collected in Recife to examine transfer across cities.
All configurations, including the baseline, were evaluated on the same held-out set of 1,151 Boston tiles.}

\myparagraph{Metrics}
\highlight{
Our repair model outputs corrected segmentation masks of sidewalk surfaces.
To directly evaluate these masks, we first report Intersection over Union and Recall (IoU and Rec) as measures of agreement with the ground truth.
Both IoU and Rec are computed on sidewalk surface areas rather than centerlines.
For each configuration we compare its predicted sidewalk pixels, \tilen's for the baseline and the repair model's otherwise, against the city's sidewalk polygons rasterized to the same grid.}
In pedestrian-network extraction, however, standard pixel-level metrics can be misleading: a prediction that correctly captures every sidewalk but is offset by a few pixels will incur a disproportionate penalty, while a prediction that hallucinates small, isolated blobs in empty tiles inflates the error count without reflecting a meaningful extraction failure. 
To address this, we introduce \textit{revised} variants (rIoU and rRec) that suppress evaluation noise from four sources: (i) disconnected components under 50 pixels, which typically arise from compression artifacts or stray classifications rather than true sidewalk geometry; (ii) tiles whose ground-truth mask is under 50 pixels, where a single misclassified pixel can swing IoU by tens of percentage points; (iii) predictions within ten pixels of the tile boundary, which penalize correct sidewalks that are split across adjacent tiles; and (iv) predicted pixels within ten pixels of a true sidewalk pixel, which absorb the small spatial offsets inherent in rasterized vector ground truth. After applying these filters we erode the remaining error-region boundaries by ten pixels to ensure that only the core disagreement area contributes to the score.
% %
% 
\highlight{In addition to these four segmentation-focused metrics, which evaluate the areas detected by the model as potentially containing sidewalks, we report four graph metrics that focus on the quality of the final sidewalk network.
The first two metrics, the average number of connected components and dangling endpoints (nodes with degree 1) per tile, compare network structure with the ground truth.
The other two, reachability and Average Path Length Similarity (APLS), compare route connectivity and lengths.
We evaluated routing within each micro tile by considering its larger context.
For each test micro tile, we considered routes within its encompassing meso tile and selected 10 random paths that started in the evaluated micro tile.
For route reachability, we computed the proportion of sampled ground-truth routes whose start and end points were also connected in the model-generated network: 1 means all sampled pairs remain connected, and 0 means none do.
We also computed APLS to compare route lengths in the ground-truth and model-generated networks.
APLS is defined as $1 - \min(1, |L_{\mathrm{GT}} - L_{\mathrm{pred}}| / L_{\mathrm{GT}})$, averaged over all sampled pairs, where $L_{\mathrm{GT}}$ and $L_{\mathrm{pred}}$ are the ground-truth and predicted network path lengths (with $L_{\mathrm{pred}}=0$ for disconnected endpoint pairs).
An APLS score of 1 indicates matching route lengths, while lower scores indicate greater disagreement.
}

\myparagraph{Validation set}
The \highlight{1,151} test tiles are stratified into four difficulty bins based on how well~\tilen's baseline segmentation recovers the ground truth, measured by rRec. 
Tiles in [0, 0.25) are the hardest, where \tilen recovers less than 25\% of the true sidewalk pixels,  while tiles in [0.75, 1.0] are the easiest, where the baseline already performs well. This stratification allows us to assess whether the repair model's targeted corrections provide the most benefit precisely where the baseline fails.

\begin{figure*}[!t]
    \centering\includegraphics[width=\linewidth]{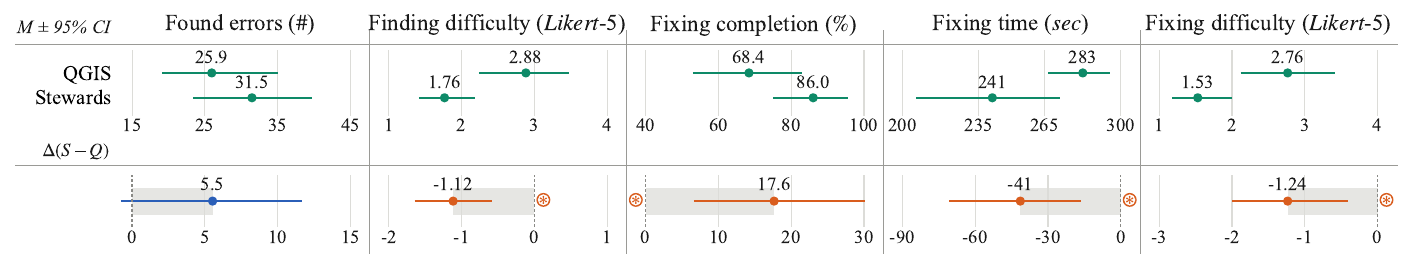}
    \vspace{-0.5cm}
  % \caption{User study results}
    \definecolor{figmean}{HTML}{0E8A68}  
    \definecolor{figdiff}{HTML}{2B5FB8}
    \definecolor{figevid}{HTML}{D95D1E}  
    \caption{\highlight{
    Results of the user study comparing QGIS and \system on finding and fixing sidewalk errors. 
    From left to right, the columns report: the number of micro tiles with errors found in two minutes, the reported difficulty of finding errors, the percentage of a micro tile fixed within the five minute limit, the time until the fix was completed, and the reported difficulty of fixing (Likert-5 scale). The top row shows the \textcolor{figmean}{means} for each tool, and the bottom row the \textcolor{figdiff}{paired differences}. 
    Error bars are 95\% bootstrap confidence intervals. Difference intervals excluding zero are
    drawn in \textcolor{figevid}{orange} and marked with $\circledast$.
    }
    }
    \label{fig:user_study1}
\end{figure*}

\highlight{
\myparagraph{Results}
Table~\ref{tab:network_evaluation} shows results across difficulty levels.
We include three repair models, repair-5, repair-10, and repair-20, trained using 5, 10, and 20 user corrections collected through the interface in Boston, the same city from which the test tiles were drawn.
We also include repair-recife-20, trained using 20 user corrections from Recife.
We focus primarily on repair-20.
Its segmentation gains are most pronounced in the two lower-recall categories, where~\tilen struggles most.
On very hard tiles ([0, 0.25)), rRec increases from 0.05 to 0.51 (+0.46), and on hard tiles ([0.25, 0.5)) from 0.39 to 0.70 (+0.31), showing that a small number of city-specific corrections can recover a large fraction of missed sidewalk coverage in these cases.
The gains diminish as tiles become easier.
On easy tiles ([0.75, 1.0]), rIoU and IoU show slight regressions, suggesting that the repair model can over-correct regions where~\tilen already performs well.
For the network metrics, repair-20 brings the average number of \textit{connected components} closer to the ground truth than~\tilen in the [0.25, 0.5) and [0.5, 0.75) categories, while slightly increasing the discrepancy in the [0, 0.25) and [0.75, 1.0] categories.
In every difficulty category, however, the average component count differs from the ground-truth average by less than one component.
The number of \textit{dangling endpoints} moves further from the ground-truth count in the three lower-recall categories, suggesting that additional network post-processing may be needed to address these remaining errors.
Repair-20 improves both routing metrics, \textit{reachability} and \textit{APLS}, across all four difficulty categories relative to the original \tilen network.
The largest gains occur on hard tiles ([0.25, 0.5)), where reachability increases from 0.23 to 0.44 (+0.21) and APLS from 0.22 to 0.40 (+0.18).
These results indicate improved routing performance in the repaired network.
Increasing the number of corrections does not improve every metric consistently, as shown by repair-5, repair-10, and repair-20.
For example, rIoU generally improves with additional corrections, consistent with more segmentation examples producing more stable predictions.
However, the routing metrics do not improve consistently as more user corrections are added.
For cross-city transfer, repair-recife-20 remains close to the original \tilen model, suggesting limited transfer from Recife to Boston and a benefit from corrections collected in the local context.
Overall, the segmentation and routing gains on very hard and hard tiles support the value of the active learning loop: a small number of city-specific corrections can improve sidewalk recovery and routing.
}

%%%%%%%%%%%%%%%%%%%%%%%%%%%%%%%%%%%%%%%%%%%%%%%%%%%%%%%%%
%%%%%%%%%%%%%%%%%%%%%%%%%%%%%%%%%%%%%%%%%%%%%%%%%%%%%%%%%
\highlight{\subsection{User study}  \label{subsec:user_study}}
%%%%%%%%%%%%%%%%%%%%%%%%%%%%%%%%%%%%%%%%%%%%%%%%%%%%%%%%%
%%%%%%%%%%%%%%%%%%%%%%%%%%%%%%%%%%%%%%%%%%%%%%%%%%%%%%%%%

\highlight{To evaluate \system in relation to conventional sidewalk-mapping workflows, we conducted a comparative user study centered on the question of whether \system enables users to audit and correct sidewalk networks more efficiently than conventional practices in urban planning and GIS. 
As a baseline, we chose QGIS, a popular GIS tool used for sidewalk network editing.
This study was approved by the University of Illinois Chicago Office for the Protection of Research Subjects [STUDY2026-0930]. 

% \myparagraph{Visualization and Interaction Conditions}

\myparagraph{Datasets} The study used data from two geographically distinct contexts: Boston, USA, and Recife, Brazil. For both locations, the dataset included orthorectified imagery together with the corresponding sidewalk network extracted by \tilen, which served as the underlying network representation for the evaluation.

\myparagraph{Participants} The sample comprised 17 graduate researchers (3 women and 14 men), including 13 PhD and 4 MS students, with backgrounds 
in urban data visualization (6), computer science (4), data visualization (3), mechanical engineering (3), and urban planning (1).
Most participants reported no prior experience with comparable GIS-based sidewalk-mapping workflows. 
All participants volunteered for the study, provided informed consent before participation, and received no monetary compensation.

\myparagraph{Procedure} The sessions were conducted in person using workstations equipped with a mouse, keyboard, and two monitors. Participants interacted with the same locally hosted study build of \system and with QGIS through a prepared project containing the corresponding spatial data. Each session began with a 1-minute introduction to the study and sidewalk data review process. Both experimental conditions were preceded by 2 minutes of tool-specific instruction and practice in areas separate from those used for evaluation. A researcher remained present throughout to provide instructions and clarify procedural questions. Following this preparation, participants completed two tasks covering complementary stages of the sidewalk-network inspection and correction workflow: one focused on identifying potential network errors (2 minutes with each tool) and the other on carrying out an end-to-end correction process (up to 5 minutes with each tool). The study concluded with post-task responses, bringing the total session duration to approximately 30 minutes.

\myparagraph{Task 1: Finding sidewalk errors} Asked to identify sidewalk-network inconsistencies, participants started from the full Dorchester study area in both conditions. 
\system presented the area through its main interface, whereas QGIS used the corresponding orthorectified imagery overlaid with the same \tilen network. 
Each participant was randomly assigned to complete the task first with either QGIS or \system.
For each system, the task comprised four 30-second rounds, each requiring participants to move to a new location and inspect an area of their choice for potential errors. 
Beyond the shared spatial data, \system exposed model uncertainty through a heatmap to support exploration.

\myparagraph{Task 2: Correcting the sidewalk network} The second task evaluated the complete sidewalk review and correction workflow within a single tile using data from Recife, Brazil. To support a controlled comparison without repeated exposure to the same network, the study used three matched tile pairs selected to balance workload across conditions. 
For each participant, randomization determined both the allocation of the matched tiles between \system and QGIS and the tool used first. 
In \system, participants started from the network produced by the repair model, whereas in QGIS they started from the original \tilen network.
Each trial lasted up to five minutes and ended earlier if no further edits were considered necessary. Participants then estimated task completion using four categories: 0-25\%, 25-50\%, 50-75\%, or 75-100\%. 
Trials that ended before the time limit were recorded as complete (100\%). For analysis, each range was converted to its midpoint.

\myparagraph{Results}
Quantitative results are reported as sample means and 95\% confidence intervals (CIs) obtained by resampling participants with replacement using 10,000 percentile bootstrap resamples, as summarized in Figure~\ref{fig:user_study1}.
Given the paired study design, condition differences are summarized by mean differences with corresponding 95\% CIs, computed within participants and defined as Stewards minus QGIS. These estimates convey both the magnitude of the observed differences and the uncertainty surrounding them. 

\myparagraph{Found errors}
In Task 1, participants marked tiles as containing potential sidewalk-network errors. They marked an average of 31.5 tiles with \system (CI = [23.4, 39.8]), compared with 25.9 using QGIS (CI = [19.2, 35.1]), yielding an estimated paired difference of 5.5 tiles in favor of \system (CI = [$-$0.8, $+$11.6]).

\myparagraph{Finding difficulty}
% Perceived difficulty showed a clearer contrast, 
Participants reported less difficulty finding errors with \system than with QGIS on a five-point Likert scale, with mean ratings of 1.76 for \system (CI = [1.41, 2.18]) and 2.88 for QGIS (CI = [2.24, 3.47]). The mean difference was $-$1.12 points (CI = [$-$1.65, $-$0.59]).

\myparagraph{Fixing completion} 
Estimated completion was higher with \system, averaging
86.0\% with \system (CI = [75.0, 95.6]), compared with 68.4\% using QGIS (CI = [52.9, 83.1]). The mean difference was $+$17.6 percentage points (CI = [$+$6.6, $+$30.1]).

\myparagraph{Fixing time} 
Participants completed the correction task faster with \system, spending an average of 241 s with \system (CI = [206, 272]), compared with 283 s using QGIS (CI = [267, 295]).
 This resulted in a mean difference of \mbox{$-$41 s} (CI = [$-$71, $-$16]).

\myparagraph{Fixing difficulty} Participants also perceived the correction process as less difficult with \system, with a mean rating of 1.53 (CI = [1.18, 2.00]) on a five-point Likert scale, compared with 2.76 for QGIS (CI = [2.12, 3.41]). The corresponding difference in perceived difficulty between the two systems was $-$1.24 points (CI = [$-$2.00, $-$0.41]).
}

\subsection{Case studies}  \label{subsec:case_studies}

The case studies exercise different facets of the system across varying analytical scales and geographic contexts: the first two trace targeted repair workflows on cities for which~\tilen extractions already exist, following the analyst's path from city-wide prioritization down through meso- and micro-scale correction; the third demonstrates the system's applicability beyond well-mapped North American cities, using Recife, Brazil, as a test of the end-to-end pipeline in a context where no location-specific training data was available.

\begin{figure}[t]
  \centering\includegraphics[width=\linewidth]{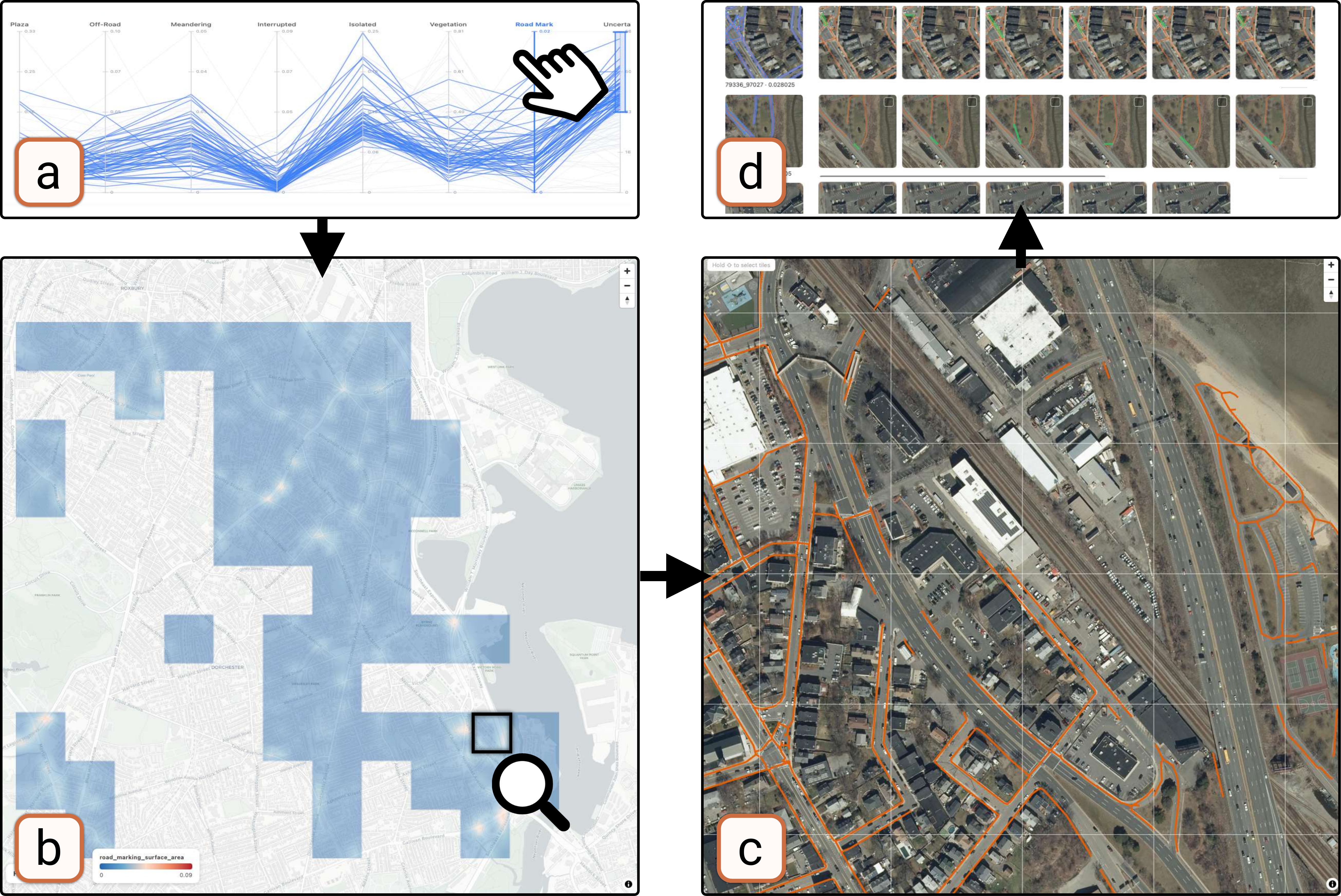}
  \caption{Starting from a city-scale uncertainty heatmap, the analyst uses the parallel coordinates plot to brush high-uncertainty tiles (a), surfacing a subset highlighted on the map (b). Gallery inspection at the meso scale reveals a residential cluster where sidewalk segments consistently fail to connect across intersections (c). The system then proposes targeted corrections for the identified dangling endpoints (d). 
}
  \label{fig:case1}
\end{figure}

\myparagraph{Case study 1: Macro-to-meso prioritized inspection}
\tilen extractions of Boston yield a network comprising 9200 tiles, of which the uncertainty heatmap flags some as high-confidence error candidates. 
Without structured prioritization, an analyst faced with this volume of suspects would have little choice but to sample randomly, an approach that mixes errors with critical connectivity failures and provides no systemic coverage guarantee. 
This case study, shown in Figure~\ref{fig:case1}, demonstrates how \system replaces that sampling regime with a data-driven triage.
The analyst begins at the macro scale, examining the uncertainty heatmap overlaid on the city map. 
Several spatial clusters of high-uncertainty tiles are immediately visible, corresponding to neighborhoods with dense tree canopy, a known correlate of segmentation failures in~\tilen's pipeline. 
Rather than drilling into the first flagged tile encountered, the analyst opens the parallel coordinates plot (a), which renders each tile as a polyline spanning the uncertainty and SAM coverage axes. 
By brushing the upper end of the uncertainty axis, the analyst filters the tile population to those most likely to contain genuine topology errors, collapsing the inspection set from thousands of candidates to a tractable subset. The tiles that satisfy the user's constraints are shown on the map (b). Reordering the PCP by SAM crosswalk coverage score further stratifies this set, surfacing tiles where high uncertainty co-occurs with substantial crosswalk presence, a combination that frequently indicates missed or misaligned intersection connections.
The tile gallery, ranked by the selected PCP dimension, provides the transition to meso scale. Scrolling through the top-ranked tiles, the analyst identifies a cluster in a residential neighborhood where the extracted network consistently fails to connect sidewalk segments across intersections (c), leaving a series of dangling endpoints on either side of each crossing.
\system's suggestions provide a few potential corrections for these endpoints (d).
The analyst proceeds to flag the affected region for repair, carrying the prioritized error set forward to the correction workflow.

\myparagraph{Case study 2: Meso-to-micro topology repair}
\begin{figure*}[t]
  \centering\includegraphics[width=\linewidth]{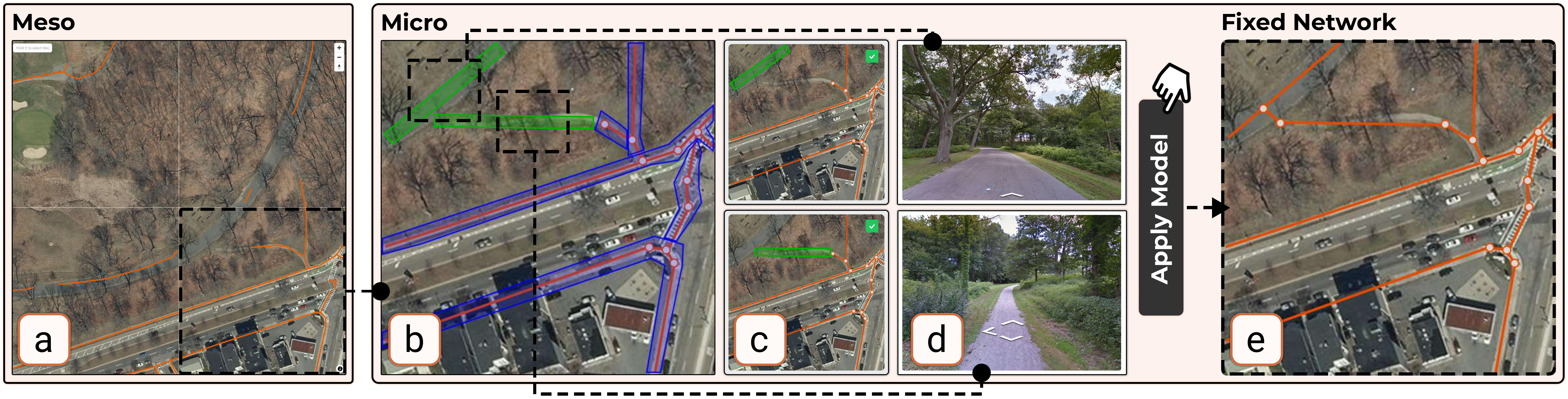}
  \vspace{-0.5cm}
  \caption{At the meso scale, dangling endpoints are highlighted as error candidates on the aerial map (a), and the system automatically proposes connecting edges for nearby endpoint pairs (b). The analyst accepts geometrically unambiguous suggestions with a single click, while ambiguous cases prompt consultation of the embedded street-level panel for ground-level verification (c, d). Where automated suggestions are insufficient, the analyst falls back to direct micro-scale editing with snapping support to hand-trace correct connections. Once a sufficient volume of corrections is accumulated, the repair model is retrained on the accepted and rejected suggestions to produce a corrected network (e), illustrating the system's intended division of labor between automated triage and human judgment.}
  \label{fig:case2}
  \vspace{-0.5cm}
\end{figure*}
Having identified a cluster of connectivity failures at the meso scale, this case study follows the repair process as the analyst descends to the micro scale and applies the system's candidate suggestions and direct editing tools to restore network continuity.
The study, shown in Figure~\ref{fig:case2}, focuses on a neighborhood in Boston with disconnected sidewalks around areas with vegetated surfaces and off-road footpaths, common errors as shown in our typology.
At the meso scale, the map view displays the extracted network overlaid on the aerial imagery, with dangling endpoints highlighted as error candidates (a).
The system has automatically generated topological repair suggestions by identifying pairs of dangling endpoints within a configurable proximity threshold and proposing connecting edges between them (b). 
These suggestions are rendered as green candidates, visually distinct from the confirmed network, allowing the analyst to evaluate each proposal in its spatial context before committing.
\highlight{For multiple flagged intersections},
the suggested connections are geometrically plausible and contextually correct: the proposed edge traces the crosswalk surface visible in the aerial imagery, and accepting it restores a clean, traversable path through the intersection. The analyst accepts these candidates with a single click per suggestion, progressively resolving the disconnected cluster.

Not all suggestions can be accepted uncritically. At one intersection in the cluster, the aerial view alone is insufficient to determine whether a proposed connection reflects a real, traversable path or an artifact of the segmentation (c). 
The analyst consults the GSV panel embedded within the interface, navigating to the flagged locations to verify ground-level conditions (d). The analyst then switches to direct micro-scale editing: zooming in until individual sidewalk centerlines are visible at the sub-meter level, snapping a new edge endpoint to the nearest confirmed node, and tracing the correct connection by hand.
The snapping behavior prevents the introduction of new near-miss gaps as the analyst draws. With a sufficient volume of corrections accumulated across the session, the analyst triggers a retraining of the repair model on the updated dataset of accepted and rejected suggestions, generating a fixed network (e). Because the model targets only~\tilen's segmentation stage, retraining replaces the upstream segmentation component while leaving the vectorization and skeletonization steps unchanged. 
The case illustrates the intended division of labor between automated suggestion and human override: the system handles the routine, geometrically unambiguous cases at speed, while the analyst's judgment, informed where necessary by street-level verification, is reserved for the contextually complex ones.

\begin{figure*}[t]
  \centering
  \includegraphics[width=\linewidth]{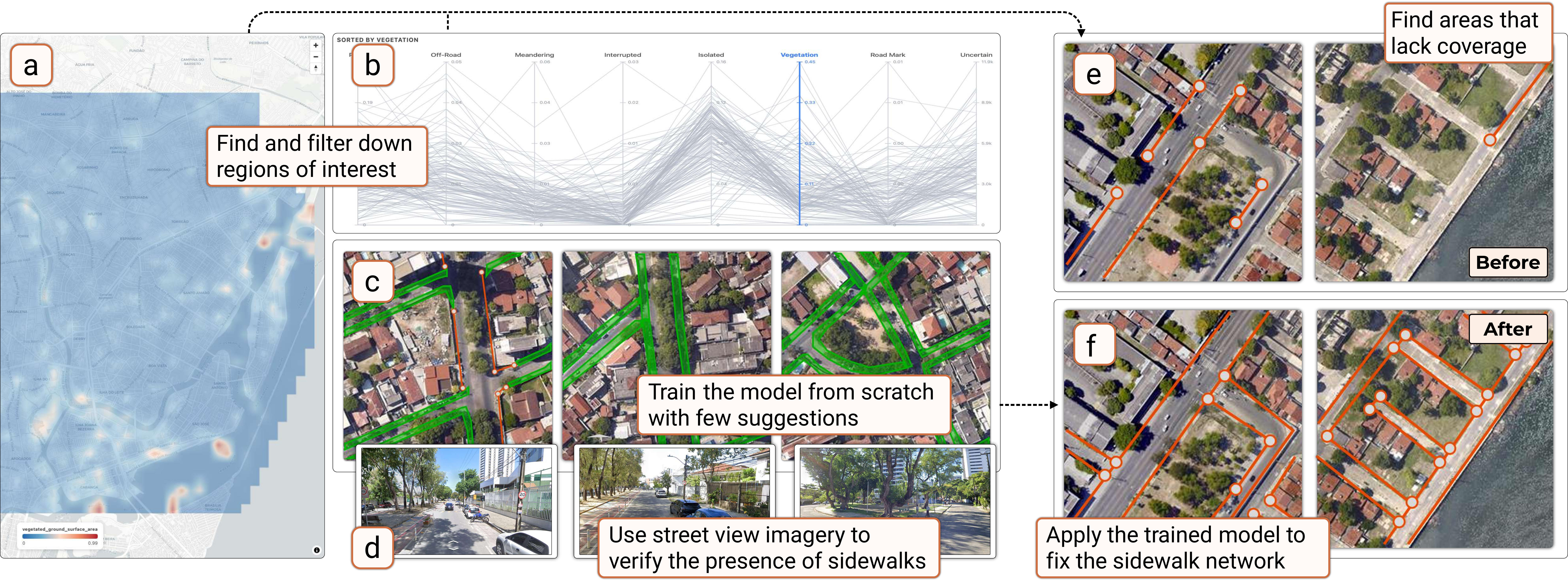}
  \caption{%
    Using \system to create, inspect, and repair the sidewalk network of Recife, a large Brazilian city with a distinct built environment. \system leverages \tilen to generate an initial version of the network. Because the city differs substantially from \tilen's training distribution, the resulting network is far from optimal. \system allows users to (a) inspect \tilen's uncertainty across the region of interest and (b) examine the distribution of features associated with typological errors. In this example, the user zooms into a specific location and (c) identifies that portions of the network are missing due to occlusion by overhead vegetation. (d) After verifying this through street-level imagery, the user (e) trains a repair model and (f) applies it to the network.
  }
  \label{fig:case3}
\end{figure*}

\myparagraph{Case study 3: Network construction for Recife}
The first two case studies demonstrate \system operating on cities with existing~\tilen extractions and relatively abundant aerial imagery of the kind the model was originally trained on. This case study asks a different question: how does the system perform when applied to a city in the Global South, where pedestrian infrastructure takes different physical forms, aerial imagery may be of lower or more variable resolution, and no location-specific training data is available to fine-tune the segmentation model? 
Recife, Brazil, a dense coastal city of approximately 1.7 million residents with a complex network of sidewalks, elevated walkways, and informal footpaths, provides a stringent test of the pipeline's generalizability. Figure~\ref{fig:case3} shows an overview of the study.

Starting from a bounding box covering Recife, the analyst runs~\tilen on publicly available aerial imagery to produce a draft pedestrian network. 
The extraction proceeds without modification to the model weights; no Recife-specific annotations are provided. 
The resulting network covers 8678 tiles, but the uncertainty heatmap reveals a substantially higher proportion of high-uncertainty tiles than was observed in the North American case studies (a), consistent with the domain shift introduced by unfamiliar pavement materials, vegetation patterns, and street geometries. 
The PCP confirms that much of this uncertainty concentrates in tiles with vegetated surfaces (b), suggesting that the segmentation model fails to correctly take into account this condition.
Rather than treating this as a failure, the analyst uses the elevated uncertainty as a structured entry point for correction. Beginning with the highest-uncertainty tiles surfaced by PCP brushing, the analyst works through the gallery and street-view imagery to assess which failures represent genuine segmentation errors versus genuine absences of infrastructure (c, d).
For tiles where sidewalk is visible in the imagery but absent from the extraction (e), direct micro-scale editing is used to trace the missing segments, with the system's snapping and suggestion mechanisms accelerating the process wherever dangling endpoints are close enough to generate candidates. 
As corrections accumulate, the system's repair model is retrained on the growing body of corrections from Recife and applied to the tiles (f).
The case demonstrates that \system is not limited to polishing already-good extractions: with iterative human correction feeding back into the model, it can serve as an active bootstrapping tool for cities where automated extraction alone cannot yet produce a usable network.

\subsection{Expert feedback} \label{subsec:expert_feedback}

To evaluate \system with domain experts, we conducted semi-structured feedback sessions with five experts who work directly on pedestrian network analysis.
E1 is a researcher specializing in urban mobility and pedestrian network modeling, with hands-on experience using automated extraction tools to generate and refine sidewalk graphs for city-scale applications. 
E2 is an urban planning researcher who had recently completed a manual sidewalk correction project, specifically addressing vegetation-induced occlusion errors, as part of a sidewalk shading index study for a major Brazilian city.
E3 is an urban planning researcher with experience in GIS mapping and livable street design.
E4 is an urban planning professor with active research in under-resourced cities and urban renewal.
E5 is a researcher in urban science, developing methodologies for sidewalk activity analysis in American cities.
All participants were shown a live demonstration of the system covering all stages of the workflow, and had the opportunity to ask questions and interact with the tool throughout the session; responses were then gathered through structured questions aligned with the system's core design goals. The sessions were conducted individually.

\myparagraph{Ingestion \& detection}
All experts validated the macro-meso-micro framework (DG3) as well-aligned with how practitioners already reason about network correction tasks. 
E1 noted that her team's workflow for refining extracted networks follows the same progression (checking overall network topology at city scale before zooming into specific windows for targeted repair) and described the structure as natural and easy to follow. 
E3 mentioned that ``I would probably spend almost all of my time in the meso and the micro views.''

\myparagraph{Inspection}
All experts affirmed the value of the parallel coordinates plot for surfacing candidate error regions (DG1).
E2 found the spatial overview useful for geographic orientation, but proposed a layer-toggling interaction, mentioning that allowing users to overlay multiple feature channels simultaneously would make the macro view considerably more actionable. 
E3 was appreciative of the typology particularly its scaffolding aspects, mentioning that ``there's so many topology errors that having starting points simplifies the work a lot (...) I feel like the psychological effect of being, like, 'here is step one'.''
E5 mentioned that \system is the only system currently capable of fixing sidewalk networks in such a human-in-the-loop approach.

\myparagraph{Repair}
The automatic repair suggestions were among the most positively received features in the feedback sessions. E1 contrasted the suggestion-first workflow with her team's prior experience of purely manual correction and estimated that starting from a pre-generated candidate would save substantial time. 
E2 highlighted an additional benefit: a suggestion already rendered on screen communicates the expected correction format implicitly, functioning as a visual prompt that reduces ambiguity about what an acceptable edit looks like.
All experts also independently praised the street-level imagery integration for ground-truth verification (DG2), describing cross-validation with images as an established part of their existing workflows that currently requires switching between separate applications.
E2 described the friction of the current multi-tab process vividly: finding the exact intersection, orienting the view to the correct cardinal direction, and reconciling two perspectives consumes meaningful time and concentration, making the embedded panel a genuine improvement.
E1 suggested organizing the repair stage around error categories, so that a user can address all vegetation-occlusion errors in a session before moving to connectivity errors, making the correction feel more structured and cognitively manageable (DG7). 
E4 mentioned that the approach presented is ``extremely pertinent, it progressively refines toward the problem.''
He further elaborated, mentioning that \system's approach can be particularly useful for cities where urbanization might lead to self-built spaces ``where the network the system can identify are not visible.''

\section{Conclusion}

\system addresses a practical bottleneck in urban accessibility work: automated pedestrian network extraction is scalable but error-prone, and correcting its outputs at scale has no established tooling.
By embedding a structured error typology into a multi-scale detect–inspect–repair workflow, the system guides analysts from city-wide triage to individual edge repairs while feeding corrections back into a retraining pipeline that improves future extractions. 
\highlight{
In the comparative user study, participants reported less difficulty finding and correcting errors with \system than with QGIS. Correction tasks showed shorter durations and higher estimated completion percentages.
Expert feedback confirms that the workflow matches how analysts reason about these errors. Our quantitative evaluation showed that a modest corpus of user corrections can improve sidewalk segmentation and routing metrics, particularly in areas where the baseline performs poorly.
}

The system's current limitations point directly to future work. Performance depends on the availability of high-resolution aerial imagery, a constraint that disproportionately affects municipalities that stand to benefit most; integrating alternative data sources, such as LiDAR, could extend applicability to underserved cities. 
\highlight{
Despite improvements in segmentation and routing, dangling-endpoint counts can move further from ground truth; additional network post-processing could address these remaining errors.  The limited transfer observed between cities suggests that corrections from the target local context remain important for adapting the repair model. The error typology, derived from Boston imagery, may not cover all failure modes encountered in broader deployment; a mechanism for analysts to propose and validate new categories would make the taxonomy more generalizable to other environments. Automatic repair suggestions connect nearby endpoints and therefore address only a subset of potential errors; they may also introduce unnecessary connections. Although the system supports city-scale inspection, correction at full city scale remains to be evaluated.
As future work, we will also investigate the integration of sunlight access models~\cite{omar2024deep} into \system as another data layer associated with errors.
}

% \fabio{\section*{Acknowledgments} }

\highlight{
\section*{Acknowledgments}
This work was supported by the U.S. National Science Foundation (NSF, Awards \#2320261, \#2330565, and \#2411223).
}

% \stefan{
% {\appendix[User study setup]
% }
% }

\bibliographystyle{IEEEtran}
% \bibliographystyle{abbrv-doi-hyperref-narrow}
%\bibliographystyle{abbrv-doi}
%\bibliographystyle{abbrv-doi-narrow}

%\nocite{*}

\bibliography{references}

% \newpage

% \section{Biography Section}

% \begin{IEEEbiographynophoto}{Stefan Cobeli}
% is currently working toward the PhD degree in computer science at the University of Illinois Chicago. His research centers on data visualization, with a focus on applying machine learning techniques for urban analytics.
% \end{IEEEbiographynophoto}

% \stefan{Bio for Leonardo, Logan,}

% \luiza{Bio for Luiza}

% \begin{IEEEbiographynophoto}{Nivan Ferreira}
% received the PhD degree in computer science from New York University. He is an assistant professor at Universidade Federal de Pernambuco. His research focuses on many aspects of interactive data visualization, in particular systems and techniques for the analysis of spatiotemporal datasets.
% \end{IEEEbiographynophoto}

% \sourav{Bio for Sourav and Peyman}

% \begin{IEEEbiographynophoto}{Fabio Miranda}
% received the PhD degree in computer science from New York University. He is an assistant professor in the Department of Computer Science at the University of Illinois Chicago. His research focuses on large-scale data analysis, data structures, visualization, and visual
% analytics.
% \end{IEEEbiographynophoto}

% \maryam{Bio for Maryam}

\vfill

\end{document}